\documentclass[trackchanges]{aastex701}
\usepackage[T5,T1]{fontenc}
\usepackage{booktabs}
\usepackage{xcolor}
\usepackage{comment}

\usepackage{caption}
\usepackage{makecell}

\usepackage{hyperref}

\begin{document}

\title{Search for Coincidences between IceCube Sub-TeV Neutrinos and Sub-Threshold Gravitational Wave Events in the LIGO-Virgo-KAGRA Third Observing Run}

\affiliation{III. Physikalisches Institut, RWTH Aachen University, D-52056 Aachen, Germany}
\affiliation{Department of Physics, University of Adelaide, Adelaide, 5005, Australia}
\affiliation{Dept. of Physics and Astronomy, University of Alaska Anchorage, 3211 Providence Dr., Anchorage, AK 99508, USA}
\affiliation{School of Physics and Center for Relativistic Astrophysics, Georgia Institute of Technology, Atlanta, GA 30332, USA}
\affiliation{Dept. of Physics, Southern University, Baton Rouge, LA 70813, USA}
\affiliation{Dept. of Physics, University of California, Berkeley, CA 94720, USA}
\affiliation{Lawrence Berkeley National Laboratory, Berkeley, CA 94720, USA}
\affiliation{Institut f{\"u}r Physik, Humboldt-Universit{\"a}t zu Berlin, D-12489 Berlin, Germany}
\affiliation{Fakult{\"a}t f{\"u}r Physik {\&} Astronomie, Ruhr-Universit{\"a}t Bochum, D-44780 Bochum, Germany}
\affiliation{Universit{\'e} Libre de Bruxelles, Science Faculty CP230, B-1050 Brussels, Belgium}
\affiliation{Vrije Universiteit Brussel (VUB), Dienst ELEM, B-1050 Brussels, Belgium}
\affiliation{Dept. of Physics, Simon Fraser University, Burnaby, BC V5A 1S6, Canada}
\affiliation{Department of Physics and Laboratory for Particle Physics and Cosmology, Harvard University, Cambridge, MA 02138, USA}
\affiliation{Dept. of Physics, Massachusetts Institute of Technology, Cambridge, MA 02139, USA}
\affiliation{Dept. of Physics and The International Center for Hadron Astrophysics, Chiba University, Chiba 263-8522, Japan}
\affiliation{Department of Physics, Loyola University Chicago, Chicago, IL 60660, USA}
\affiliation{Dept. of Physics and Astronomy, University of Canterbury, Private Bag 4800, Christchurch, New Zealand}
\affiliation{Dept. of Physics, University of Maryland, College Park, MD 20742, USA}
\affiliation{Dept. of Astronomy, Ohio State University, Columbus, OH 43210, USA}
\affiliation{Dept. of Physics and Center for Cosmology and Astro-Particle Physics, Ohio State University, Columbus, OH 43210, USA}
\affiliation{Niels Bohr Institute, University of Copenhagen, DK-2100 Copenhagen, Denmark}
\affiliation{Dept. of Physics, TU Dortmund University, D-44221 Dortmund, Germany}
\affiliation{Dept. of Physics and Astronomy, Michigan State University, East Lansing, MI 48824, USA}
\affiliation{Dept. of Physics, University of Alberta, Edmonton, Alberta, T6G 2E1, Canada}
\affiliation{Erlangen Centre for Astroparticle Physics, Friedrich-Alexander-Universit{\"a}t Erlangen-N{\"u}rnberg, D-91058 Erlangen, Germany}
\affiliation{Physik-department, Technische Universit{\"a}t M{\"u}nchen, D-85748 Garching, Germany}
\affiliation{D{\'e}partement de physique nucl{\'e}aire et corpusculaire, Universit{\'e} de Gen{\`e}ve, CH-1211 Gen{\`e}ve, Switzerland}
\affiliation{Dept. of Physics and Astronomy, University of Gent, B-9000 Gent, Belgium}
\affiliation{Dept. of Physics and Astronomy, University of California, Irvine, CA 92697, USA}
\affiliation{Karlsruhe Institute of Technology, Institute for Astroparticle Physics, D-76021 Karlsruhe, Germany}
\affiliation{Karlsruhe Institute of Technology, Institute of Experimental Particle Physics, D-76021 Karlsruhe, Germany}
\affiliation{Dept. of Physics, Engineering Physics, and Astronomy, Queen's University, Kingston, ON K7L 3N6, Canada}
\affiliation{Department of Physics {\&} Astronomy, University of Nevada, Las Vegas, NV 89154, USA}
\affiliation{Nevada Center for Astrophysics, University of Nevada, Las Vegas, NV 89154, USA}
\affiliation{Dept. of Physics and Astronomy, University of Kansas, Lawrence, KS 66045, USA}
\affiliation{UCLouvain, Centre for Cosmology, Particle Physics and Phenomenology, CP3, Chemin du Cyclotron 2, 1348 Louvain-la-Neuve, Belgium}
\affiliation{Department of Physics, Mercer University, Macon, GA 31207-0001, USA}
\affiliation{Dept. of Astronomy, University of Wisconsin{\textemdash}Madison, Madison, WI 53706, USA}
\affiliation{Dept. of Physics and Wisconsin IceCube Particle Astrophysics Center, University of Wisconsin{\textemdash}Madison, Madison, WI 53706, USA}
\affiliation{Institute of Physics, University of Mainz, Staudinger Weg 7, D-55099 Mainz, Germany}
\affiliation{Department of Physics, Marquette University, Milwaukee, WI 53201, USA}
\affiliation{Institut f{\"u}r Kernphysik, Universit{\"a}t M{\"u}nster, D-48149 M{\"u}nster, Germany}
\affiliation{Bartol Research Institute and Dept. of Physics and Astronomy, University of Delaware, Newark, DE 19716, USA}
\affiliation{Dept. of Physics, Yale University, New Haven, CT 06520, USA}
\affiliation{Columbia Astrophysics and Nevis Laboratories, Columbia University, New York, NY 10027, USA}
\affiliation{Dept. of Physics, University of Oxford, Parks Road, Oxford OX1 3PU, United Kingdom}
\affiliation{Dipartimento di Fisica e Astronomia Galileo Galilei, Universit{\`a} Degli Studi di Padova, I-35122 Padova PD, Italy}
\affiliation{Dept. of Physics, Drexel University, 3141 Chestnut Street, Philadelphia, PA 19104, USA}
\affiliation{Physics Department, South Dakota School of Mines and Technology, Rapid City, SD 57701, USA}
\affiliation{Dept. of Physics, University of Wisconsin, River Falls, WI 54022, USA}
\affiliation{Dept. of Physics and Astronomy, University of Rochester, Rochester, NY 14627, USA}
\affiliation{Department of Physics and Astronomy, University of Utah, Salt Lake City, UT 84112, USA}
\affiliation{Dept. of Physics, Chung-Ang University, Seoul 06974, Republic of Korea}
\affiliation{Oskar Klein Centre and Dept. of Physics, Stockholm University, SE-10691 Stockholm, Sweden}
\affiliation{Dept. of Physics and Astronomy, Stony Brook University, Stony Brook, NY 11794-3800, USA}
\affiliation{Dept. of Physics, Sungkyunkwan University, Suwon 16419, Republic of Korea}
\affiliation{Institute of Physics, Academia Sinica, Taipei, 11529, Taiwan}
\affiliation{Dept. of Physics and Astronomy, University of Alabama, Tuscaloosa, AL 35487, USA}
\affiliation{Dept. of Astronomy and Astrophysics, Pennsylvania State University, University Park, PA 16802, USA}
\affiliation{Dept. of Physics, Pennsylvania State University, University Park, PA 16802, USA}
\affiliation{Dept. of Physics and Astronomy, Uppsala University, Box 516, SE-75120 Uppsala, Sweden}
\affiliation{Dept. of Physics, University of Wuppertal, D-42119 Wuppertal, Germany}
\affiliation{Deutsches Elektronen-Synchrotron DESY, Platanenallee 6, D-15738 Zeuthen, Germany}

\author[0000-0001-6141-4205]{R. Abbasi}
\affiliation{Department of Physics, Loyola University Chicago, Chicago, IL 60660, USA}
\email{rabbasi@luc.edu}

\author[0000-0001-8952-588X]{M. Ackermann}
\affiliation{Deutsches Elektronen-Synchrotron DESY, Platanenallee 6, D-15738 Zeuthen, Germany}
\email{markus.ackermann@desy.de}

\author{J. Adams}
\affiliation{Dept. of Physics and Astronomy, University of Canterbury, Private Bag 4800, Christchurch, New Zealand}
\email{jenni.adams@canterbury.ac.nz}

\author[0000-0003-2252-9514]{J. A. Aguilar}
\affiliation{Universit{\'e} Libre de Bruxelles, Science Faculty CP230, B-1050 Brussels, Belgium}
\email{juanan.aguilar@icecube.wisc.edu}

\author[0000-0003-0709-5631]{M. Ahlers}
\affiliation{Niels Bohr Institute, University of Copenhagen, DK-2100 Copenhagen, Denmark}
\email{mahlers@icecube.wisc.edu}

\author[0009-0001-2444-4162]{S. Ali}
\affiliation{Dept. of Physics and Astronomy, University of Kansas, Lawrence, KS 66045, USA}
\email{shoukat@ku.edu}

\author{N. M. Amin}
\affiliation{Bartol Research Institute and Dept. of Physics and Astronomy, University of Delaware, Newark, DE 19716, USA}
\email{moureen@udel.edu}

\author[0000-0001-9394-0007]{K. Andeen}
\affiliation{Department of Physics, Marquette University, Milwaukee, WI 53201, USA}
\email{karen.andeen@icecube.wisc.edu}

\author[0000-0003-4186-4182]{C. Arg{\"u}elles}
\affiliation{Department of Physics and Laboratory for Particle Physics and Cosmology, Harvard University, Cambridge, MA 02138, USA}
\email{carlos.arguelles@icecube.wisc.edu}

\author{S. Athanasiadou}
\affiliation{Deutsches Elektronen-Synchrotron DESY, Platanenallee 6, D-15738 Zeuthen, Germany}
\email{sofia.athanasiadou@icecube.wisc.edu}

% ...

\date{\today}

\collaboration{all}{IceCube Collaboration}

%\author[orcid=0000-0002-0962-4878,sname='Mukherjee']{Tista Mukherjee}
%\affiliation{Institute for Astroparticle Physics, Karlsruhe Institute of Technology}
%\author{The IceCube Collaboration}

% Your emails:
%\emailAdd{tista.mukherjee@kit.edu}
%\emailAdd{second.author@inst.edu}
\email[show]{analysis@icecube.wisc.edu}

\begin{abstract}

Since the first direct detection of gravitational waves (GW) by the LIGO-Virgo-KAGRA detectors, the IceCube Neutrino Observatory has searched for their possible neutrino counterparts. The sub-threshold GW alerts from the third observation run of LIGO-Virgo-KAGRA, identified by both template-based searches and minimally modelled pipelines, have also been examined in follow-up analyses with TeV–PeV neutrinos, but no correlations were found. This work presents a systematic search for $\mathcal{O}(10-100)$ GeV neutrino emission from the sub-threshold GWs, utilising the archival all-flavour, sub-TeV neutrino sample from the IceCube Neutrino Observatory. From the public GW catalogues, we select 103 candidates with $p_\mathrm{astro} \geq $ 0.1, comprising a mixture of plausible source classes of compact binary mergers and GW burst-like triggers. A catalogue search is then conducted to identify correlated sub-TeV neutrinos within a 1000~s time window, employing an unbinned maximum-likelihood method to assess the statistical significance of spatial and temporal coincidences. The most significant correlation, with a $p$-value of 0.04\% before trial correction, was associated with a binary black hole merger. However, after trial correction, no statistically significant coincidences were found. We report upper limits to the flux of neutrinos from the sub-threshold GW sources followed up in this analysis, which lie in the range of 1 - 10 GeV $\mathrm{cm}^{-2}$ at the $\mathcal{O}(10-100)$ GeV neutrino energy range.

\end{abstract}

%\keywords{\uat{Multi-messenger searches}{573} --- \uat{Neutrinos}{343} --- \uat{Gravitational Waves}{739} --- \uat{IceCube}{847} --- \uat{Archival searches}{} --- \uat{Solar physics}{}}

\section{Introduction}

The most energetic phenomena in the high-energy, non-thermal Universe remain among the most puzzling and least understood to date.
%For over a century, CRs (CRs) and photons in a broad band of wavelengths have been considered as our primary messengers from such high-energy, non-thermal astrophysical sources. However, both come with inherent fundamental limitations, as photons are absorbed and CR get deflected during their journey towards Earth.
To better understand them, cosmic messengers that travel vast distances without alteration can serve as a valuable source of information. Astrophysical neutrinos are well suited for this purpose because they are produced in hadronic interactions of cosmic rays (CR) at their sources.
%alongside cosmic rays (CR) via hadronic interactions in similar astrophysical environments.
As neutral, weakly interacting particles, they arrive at Earth without interaction or deflection, preserving directional and spectral information about their sources, while carrying unique information about the physical processes occurring there.
In 2013, IceCube, a neutrino observatory in Antarctica, reported 28 high-energy neutrino event candidates in the energy range $30-1200$~TeV \citep{IC_AstroNu_2013}, revealing a diffuse high-energy neutrino flux of extragalactic origin. This opened a new window to the high-energy Universe. Two years after that, the LIGO detector directly observed gravitational waves (GW) from a binary black hole (BBH) merger. The event occurred $\sim\,410$ Mpc from Earth \citep{GW150914}, marking the beginning of a new era of observational astronomy. Combining the electromagnetic (EM), neutrino, and GW observations allows us to study cosmic sources in the non-thermal Universe via multi-messenger follow-up searches. Even the absence of coincident signals provides meaningful constraints, contributing to our understanding of the most energetic and exotic environments of the Universe.

Since the advent of multi-messenger studies, coincident detections of GWs and EM radiation have been made \citep{GW170817}, confirming the association of short gamma-ray bursts and the subsequent kilonovae with the merger of compact binary systems, specifically neutron-star binaries.
%This was the first-ever detection of a multi-messenger source.
At the same time, evidence of joint sources of EM radiation and neutrinos continues to emerge \citep{TXS_IC, NGC1068_IC, GP_IC}, pointing to potential hadronic interaction sites in the vicinity of CR accelerators. To obtain a complete picture, a common source of GWs and neutrinos needs to be identified. Several promising candidates, such as compact binary coalescences (CBC) with neutron star components (e.g., neutron star-black hole (NSBH) binaries and binary neutron stars (BNS)), are currently under investigation. Such a detection would open a unique window into the innermost, densest regions of hadronic accelerators, which are otherwise inaccessible to photons.
%Among the CBC sources, BNS mergers are well-established as prime multi-messenger sources, especially after the landmark GW170817 event, which had clear electromagnetic (EM) counterparts. In contrast,
%BBH mergers have not yet been associated with confirmed EM or neutrino counterparts. While they are typically
BBH mergers are generally not expected to produce EM or neutrino counterparts on their own. However, BBH systems embedded in active galactic nucleus (AGN) disks provide a compelling multi-messenger scenario. Such a source can produce photons and neutrinos through hadronic interactions with the disk gas and ambient photon fields \citep{Biswa_AGNdisk_BBH_2025, Bartos_BBH_AGN, BBH_AGN_APJL, BBH_AGN_MNRAS}.
%A BBH merger in an AGN disk may launch a relativistic jet and produce a gamma-ray burst \citep{Biswa_AGNdisk_BBH_2025}. The merger remnant can also experience enhanced accretion and high spin, powering jets that interact with the dense surrounding disk. These interactions generate shocks that accelerate CR protons, which can produce photons and neutrinos through hadronic interactions with the disk gas and ambient photon fields \citep{BBH_AGN_Antonini, Bartos_BBH_AGN, BBH_AGN_Ford, BBH_AGN_APJL, BBH_AGN_MNRAS}.
This scenario gained further interest following the observation of the optical transient ZTF19abanrhr in AGN J124942.3+344929, proposed as a possible EM counterpart to GW190521 \citep{ZTF_BBH_kilonova}, detected by the LIGO-Virgo-KAGRA (LVK) network during the first half of the third observation run (O3a) \citep{GW190521}.

Despite all these promising scenarios, the `smoking gun' signature of high-energy neutrino observation from any of the observed compact binary mergers is yet to be detected.
%Nevertheless, CBC has already been established as an interesting multi-messenger source, presenting multiple promising avenues for CR acceleration and connecting GW, CR, photons, and neutrinos within a unified framework.
%To address this challenge, the IceCube Neutrino Observatory is playing an active role, both in real-time and archival follow-ups of high-significance GW alerts issued by the LVK collaboration
Extensive real-time and archival searches by the IceCube Neutrino Observatory have investigated neutrino emission associated with high-significance GW alerts across a broad range of time windows, to look for neutrino counterparts within a wide range of $\rm MeV-\rm PeV$ energies, considering short- and long-duration time windows~\citep{GWHENsearch_GW170817, jointGWHENsearch_O1, GWTC1_ICfollowup, GWfollowupO3, greco-gw, greco_Erratum, gw_nu, ICRC25_aswathi, GFU_O4follow_up, ELOWEN_LVKO4followUp, LVKO4a_realTime_GFU, GWfollowUp_MeVnu}. The Pierre Auger Observatory, Baikal-GVD, ANTARES, KM3NeT, and Super-KamiokaNDE have also independently searched for neutrino counterparts to individual GW events \citep{GWHENsearch_GW170817, Auger_UHEnu_BBHmerger, ANTARES_GWfollowUp_O3, KM3NeT_O3followUp, SK_O4_GWfollowUp, GW170817_BaikalGVD}. However, the absence of confirmed joint GW-neutrino detections from these targeted follow-ups underscores the need for more comprehensive searches.

This paper addresses that need by presenting a systematic study of an extensive set of archival GW events, known as the `sub-threshold GW', detected during the LVK O3 between April 2019 and March 2020. The dataset has been publicly available since 2021. Since these GW events met a less stringent detection criterion than the highly significant events, they are classified as lower-significance GW events. Such alerts are also being publicly circulated during the fourth observation run of LVK and are followed up in real time to identify their $\rm TeV-\rm PeV$ neutrino counterparts~\citep{LVKO4a_realTime_GFU, GFU_O4follow_up}. In case of a coincident neutrino detection, it would significantly boost their astrophysical relevance \citep{elevated_pastro}.
%A coincident detection with neutrinos will boost their astrophysical relevance
The sub-threshold GWs from LVK O3 have already been considered for an archival follow-up search to identify possible $\rm TeV-\rm PeV$ neutrino counterparts for individual GW events \citep{Veske_subthGW_GFU}, but no significant correlation was confirmed.
%Previous searches have investigated possible $\rm TeV-\rm PeV$ neutrino counterparts to individual O3 sub-threshold GW candidates \citep{Veske_subthGW_GFU}, but no significant correlation was found.
%To maximise sensitivity to such counterparts, 
To further extend the scope of such a search in the frame of this work, IceCube's $\mathcal{O}(10-100)$ GeV neutrino data sample is employed. Several potential joint GW-neutrino sources are expected to produce off-axis relativistic jets.
%or have quasi-thermal neutrino emission from sub-photospheric proton–neutron interactions.
In these scenarios, the observed neutrino emission is Doppler-deboosted to the sub-TeV energies \citep{Biehl_offAxisGRB}.

%In these scenarios, the higher energy neutrino flux detectable at Earth gets Doppler de-boosted,
%to sub-TeV energies~\cite{Biehl_offAxisGRB}, explaining the non-observation of high-energy neutrino counterparts.
%yielding neutrino flux in the sub-TeV energy range on Earth due to Doppler de-boosting effects

With this motivation, the remainder of this paper is organised as follows. In Sec. \ref{sec:nu_data}, we describe the IceCube Neutrino Observatory, the neutrino detector from which we obtain the $\mathcal{O}(10-100)$ GeV neutrino data. The GW datasets containing the sub-threshold event information from O3 are discussed in Sec. \ref{sec:GW_data}. We describe the analysis method in Sec. \ref{sec:UML} and the obtained results in Sec. \ref{sec:results}. Finally, we summarise the conclusions and outlook in Sec. \ref{sec:summary}.

\section{The IceCube Neutrino Observatory and sub-TeV neutrino sample}
\label{sec:nu_data}

The IceCube Neutrino Observatory is a cubic-kilometre detector located under the Antarctic ice \citep{IC_JINST}.
%To capture the neutrino detection signature in-ice, the primary detector component comprises 86 strings containing 5160 digital optical modules (DOM) in total drilled into the glacial ice. The instrumentation begins at a depth of 1450 m. It extends to 2450 m, where the ice is sufficiently transparent to capture Cherenkov radiation emitted by the charged lepton produced by the neutrino interaction in the surrounding medium. optimally. 78 out of the 86 strings are arranged in a hexagonal pattern with an average spacing of 125 m.
To capture the in-ice neutrino detection signature, a total of 5,160 digital optical modules (DOMs) were embedded in the glacial ice at the South Pole. This detector array, completed in 2010, extends between 1450\,m and 2450\,m below the surface. At these depths, the ice is sufficiently transparent to detect Cherenkov radiation from charged leptons produced in neutrino interactions in the surrounding medium. Of the 86 strings, 78 are arranged in a hexagonal pattern with an average spacing of 125 m. The DOMs on these strings are spaced 17 m apart vertically. This configuration is designed to detect neutrinos with energies above 100 GeV from distant astrophysical sources.
%As the high-energy neutrinos typically light up a large volume in the detector, a sparse configuration allows for a cost-effective setup.

In addition, there is a denser sub-array of eight strings located at the centre of the main in-ice detector, known as `IceCube DeepCore' \citep{deepcore}. It contains seven standard IceCube strings within its fiducial region. The DOMs are placed between depths of 2100\,m and 2450\,m, in the clearest ice. Above this volume, optical scattering and absorption are significantly higher. This region is known as the dust layer. The DOMs are spaced 7\,m apart vertically, and strings are separated by an average of 72\,m horizontally. Above the dust layer, a subset of 10 DOMs is deployed within the depths of $1750-1850$\,m, with 10\,m spacing, forming a `veto cap'. It can identify and reject atmospheric muons, which significantly reduces the background rate for sub-TeV neutrinos.
%especially the ones coming vertically.
%The strings are more closely spaced within the DeepCore region, with a mean spacing of 70~m - 72~m. The vertical spacing between DOMs on a string is also smaller, at approximately 7 m.
%The spacing between the DOMs is also less here, which is approximately 7~m. 
DeepCore DOMs also feature a newer PMT model with a higher quantum efficiency than that used in the standard IceCube strings. All these features together help DeepCore lower the neutrino detection threshold to approximately 10 GeV.
%These reconstructed sub-TeV neutrino events constitute the `GRECO Astronomy' (hereafter referred to as GRECO) dataset. `GRECO' is an acronym which stands for GeV Reconstructed Events with Containment for Oscillations.
%These reconstructed sub-TeV neutrino events constitute the `GRECO Astronomy’ dataset 
The subset of reconstructed $\mathcal{O}(10-100)$ GeV neutrino events that pass the GRECO event selection criteria constitutes the `GRECO Astronomy' dataset. Hereafter, in this paper, the dataset is referred to as `GRECO', an acronym for GeV Reconstructed Events with Containment for Oscillations. Originally designed for oscillation studies, it was further optimised for transient follow-up searches \citep{GRECO_3yr_transient, IC_GRB_2022, novae, greco-gw, greco_Erratum, greco_grb}. In this work, we use the GRECO events recorded during the data-taking runs from 2012 to 2019, between 26$^\mathrm{th}$ April, 2012, and 29$^\mathrm{th}$ May, 2020.
%In contrast to IceCube's high-energy samples ($\sim \mathcal{O}(10 - 100)$~TeV), GRECO is particularly valuable for identifying lower-energy neutrino counterparts to astrophysical transients (\cite{GRECO_3yr_transient, IC_GRB_2022, greco-gw, greco_Erratum, novae, greco_grb}).
%This work focuses on the sub-TeV neutrinos detected by IceCube’s dense infill array, DeepCore (\cite{deepcore}).

%The GRECO dataset covers a reconstructed energy range of $\mathcal{O}(10 - 100)$~GeV.
%One of the most important features of GRECO is that it is an all-flavour dataset. If the parent neutrino is a muon-neutrino, the secondary muon produced via charged-current interaction can travel long distances before it further loses its energy via radiative processes like Bremsstrahlung or ionisation. As these muons scatter less in ice, they appear to have travelled along long, straight paths, which are referred to as `tracks'. For an electron neutrino, the resultant electron produces an electromagnetic shower in the local detector volume. A similar effect could be obtained from the charged current interaction of tau neutrinos and the neutral current interaction of all flavours. These are labelled as `cascade' events. Due to the inclusion of cascades with track-like events in the dataset, GRECO maintains a fairly uniform effective area across all sky regions, including the Southern Hemisphere, the Northern Hemisphere, and the celestial horizon.
A key feature of the GRECO dataset is its sensitivity to all neutrino flavours.
%Muon neutrinos undergoing charged-current interactions produce %muons that can travel long distances in ice, leaving well-defined, track-like signatures due to their relatively small scattering. In contrast, electron and tau neutrino charged-current interactions, along with neutral-current interactions of all flavours, produce electromagnetic showers, creating more spherical, cascade-like signatures.
%electron neutrinos generate EM showers that deposit energy locally within the detector volume. Similar cascade-like signatures arise from charged-current interactions of tau neutrinos and neutral-current interactions of all flavours.
Muon neutrinos undergoing charged-current interactions produce muons that can travel long distances through ice, leaving well-defined, track-like signatures.
%Due to their relatively small scattering, the tracks allow good angular reconstruction.
In contrast, charged-current interactions of electron and tau neutrinos, along with neutral-current interactions of all flavours, produce electromagnetic and hadronic showers, creating more spherical, cascade-like signatures.
%with poorer angular resolution than tracks.
%typically produce cascade-like signatures, originating from electromagnetic and hadronic showers.
Including both topologies, GRECO achieves a relatively uniform effective area across the entire sky, including the Northern and Southern Hemispheres and the celestial horizon.
%has an enhanced effective area coverage in the Southern Hemisphere. It
This contrasts with the effective area coverage of high-energy, TeV$-$PeV tracks, which is enhanced in the Northern Hemisphere but reduced in the Southern Hemisphere by selection criteria required to suppress the atmospheric muon background. For short-duration transients, sensitivity is driven by effective area. The all-sky coverage is therefore well suited for follow-up observations of GW events, which often have broad localisation regions spanning both hemispheres. %GRECO also has a relatively stable event rate, which provides an estimate of the number of background events expected across the entire sky within a given time window. The average expectation is $\sim$4.6 events in 1000~s.

GRECO exhibits a relatively stable event rate, averaging $\sim\,4.6$ events per 1000\,s. Roughly 60\% of the background comes from atmospheric neutrinos, and the rest from atmospheric muons. This provides a robust background estimate for transient searches across the entire sky \citep{greco-gw, greco_Erratum}. Seasonal variation in the atmospheric neutrino flux also affects the background event rate by $\simeq 4-10$\%. Although the background is high, it is strongly suppressed in searches for short transients. 
%The GRECO sample initially excluded an angular error estimate because it is computationally expensive. The angular uncertainty estimators, such as the Paraboloid (\cite{sigma_calc}), which rely on sampling the reconstruction algorithm multiple times across different regions of the parameter space, cannot be used as part of it.
%The angular uncertainty for the GRECO sample was estimated separately and added as part of the GRECO Astronomy. It is derived from a random forest algorithm trained on both charged-current and neutral-current interactions for neutrinos of all flavours (\cite{novae}). From the distribution of angular uncertainty with respect to reconstructed energy, it is observed that some higher-energy events may achieve resolutions as fine as 1$^\circ$. However, typical cascade-like events have uncertainties of $\sim \mathcal{O}$(10$^\circ$).
The angular uncertainty for the GRECO sample is estimated separately and included in the GRECO Astronomy dataset, based on a random forest trained on all interaction types \citep{novae}. While some high-energy tracks reach angular resolutions of $\mathcal{O}(1^\circ)$, typical cascade-like events have uncertainties of $\mathcal{O}(10^\circ)$, which limits point-source sensitivity.
Also, for low-energy muon events, the uncertainty can be large, as less energy is deposited, leading to fewer hit DOMs, which makes the event reconstruction challenging. 
%The events exhibit poorer angular resolution compared to high-energy tracks. 
%Point-source follow-up analysis using such events with large angular uncertainty essentially reduces to a counting experiment.
%The coarse angular resolution of the GRECO sample limits point-source sensitivity. 
%effectively reducing them to counting experiments. 
However, this coarse angular resolution is less critical for GW follow-up, where the source localisation regions are comparably broader.
%this coarse angular resolution is less critical for very extended and transient sources, such as GW events, where GRECO can still provide useful constraints on source locations.
%The related skymaps have much worse localisation, with a median 90\% containment area > 2000 $\mathrm{deg}^2$ (see Fig.~\ref{fig:greco_area}), which is most of the time an order more compared to the GRECO area containment, as seen in Fig~\ref{fig:greco_area}. Therefore, despite its coarse angular resolution, GRECO can help constrain the locations of GW sources.

\section{Sub-threshold Gravitational Wave Events from LIGO-Virgo-KAGRA}
\label{sec:GW_data}

%IceCube actively participates in both real-time and archival follow-ups to GW alerts issued by LVK.
The LVK public event catalogues, Gravitational Wave Transient Catalogues (GWTC)-1, 2.1, 3 \citep{gwtc-1, gwtc-2, gwtc-3} provide publicly available lists of GW candidates identified during the first three LVK observing runs. They consist of events detected with high confidence, with a False Alarm Rate (FAR) $<$ 2 $\mathrm{yr^{-1}}$.
%depending on background activity around the GW signal detection time.
They typically have high posterior probabilities of originating from CBC-like sources. LVK quantifies this by $p_\mathrm{astro}$, which is greater than 50\%.
%i.e., $p_\mathrm{astro} = \mathrm{p_{BBH} + p_{BNS} + p_{NSBH}} \geq$ 0.5 for these events. Here, $\mathrm{p_{BBH}, p_{BNS}}$, and $\mathrm{p_{NSBH}}$ are the posterior probabilities of the GW event being a BBH, BNS, or NSBH merger, respectively.
%Such events are found in the existing Gravitational Wave Transient Catalogues (GWTC) -1, 2.1, 3 \citep{gwtc-1, gwtc-2, gwtc-3}, respectively.

In addition to high-confidence detections, GWTC-2.1 contains details of 1201 low-significance or sub-threshold events detected during LVK O3a.
%They were identified by searches for CBCs, using GstLAL \citep{gstlal}, MBTA \citep{mbta}, PyCBC \citep{pycbc}, and PyCBC-Highmass \citep{pycbc-HM} detection pipelines. GWTC-3 additionally provides information on 1048 sub-threshold GW events from rest of LVK O3, including the ones identified by the coherent WaveBurst (cWB) search pipeline for unmodelled GW transients \citep{cwb_allsky} during O3. They are referred to as `cWB' events throughout this paper.
%Following O3a, GWTC-2.1 contains 1201 sub-threshold candidates, while GWTC-3 contains a further 1048 candidates from O3, including cWB candidates. We use these catalogues as the starting sample for our archival neutrino search and apply the selection criteria described below to identify the events included in our analysis.
%After the completion of O3a, a list of 1201 sub-threshold alerts was included in the GWTC-2.1 catalogue.
They were identified by four analysis pipelines, GstLAL \citep{gstlal}, MBTA \citep{mbta}, PyCBC \citep{pycbc}, and PyCBC-Highmass \citep{pycbc-HM}, all tailored to detect GW signals from CBCs.
%such as binary black holes (BBH), binary neutron stars (BNS), and neutron star-black hole (NSBH) binaries. These CBC search pipelines are named GstLAL (\cite{gstlal}), MBTA (\cite{mbta}), PyCBC (\cite{pycbc}), and PyCBC-Highmass (\cite{pycbc-HM}).
Similarly, 1048 sub-threshold GWs were further included in GWTC-3. This included not only the GW signal associated with CBC from the rest of O3, but also the CBC-optimised version of the `Coherent Wave Burst' (cWB) search pipeline \citep{cwb_allsky}, to identify GW burst-like events detected during the entire O3. The events obtained from this pipeline are referred to as `cWB' in this paper.
%Unlike the LVK GW events satisfying the O3 high-significance criteria,
All these sub-threshold GWs have $\mathrm{p_{astro}} <$ 0.5, and are characterised by $2~\mathrm{yr^{-1}} \leq \mathrm{FAR} < 2~\mathrm{day^{-1}}$.
%The quantity $p_{\rm astro}$ is the Bayesian probability of detecting a GW signal coming from a CBC-like source. Mathematically, we can express it as, $p_{\rm astro}$ = $\mathrm{p_{BBH} + p_{BNS} + p_{NSBH}}$. Here, $\mathrm{p_{BBH}, p_{BNS}}$, and $\mathrm{p_{NSBH}}$ are the probabilities for detecting the GW signal from a BBH, BNS, and NSBH-like source, respectively.
%Interestingly, GWTC-2.1 and GWTC-3 include all CBC and cWB candidates with FAR $<$ 2 $\mathrm{day^{-1}}$ ($\simeq$ 730 $\mathrm{yr^{-1}}$). This threshold encompasses both low-significance and high-significance events.

\begin{figure}[htb!]
    \centering
    \includegraphics[width=0.50\textwidth]{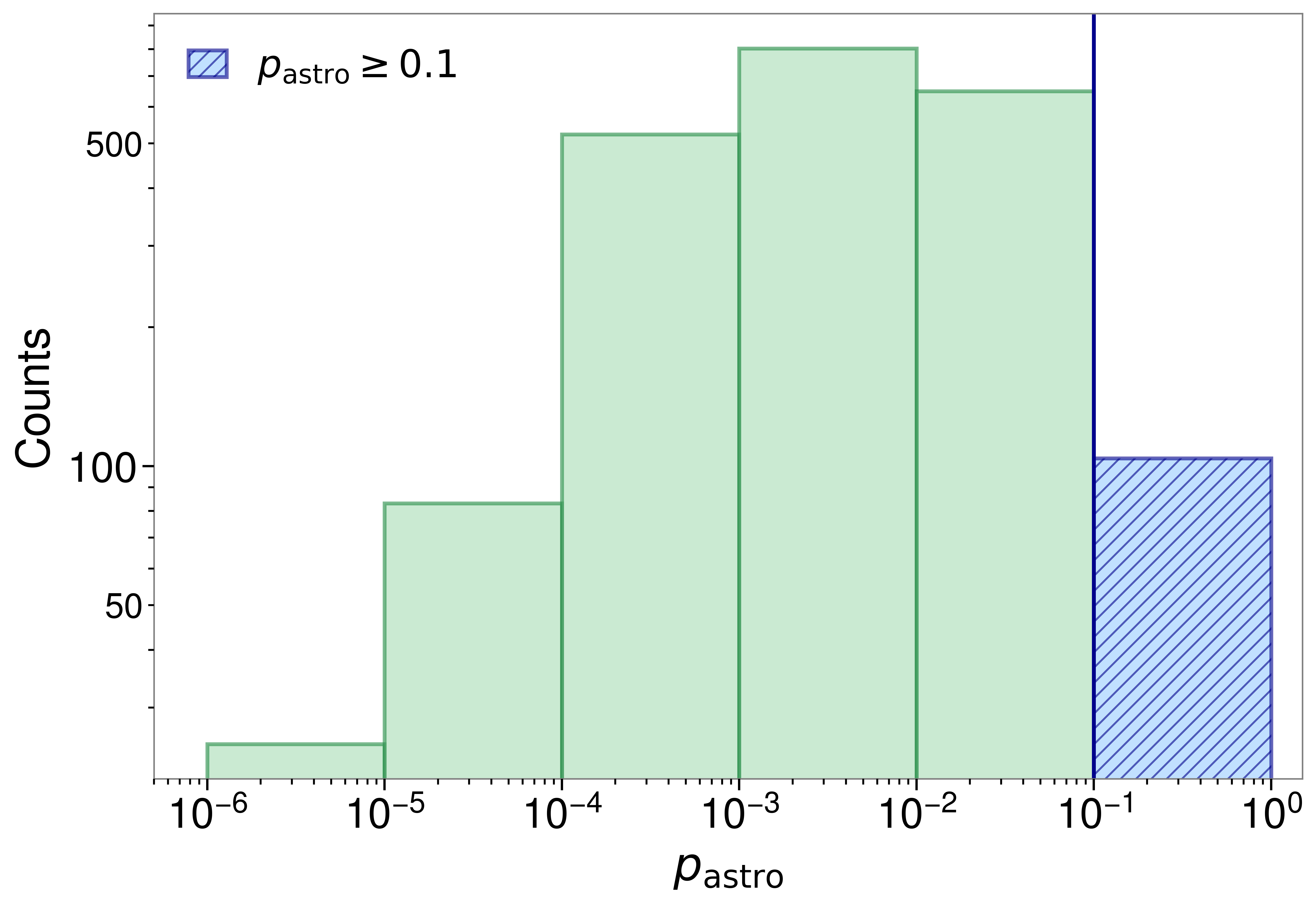}
    \includegraphics[width=0.48\textwidth]{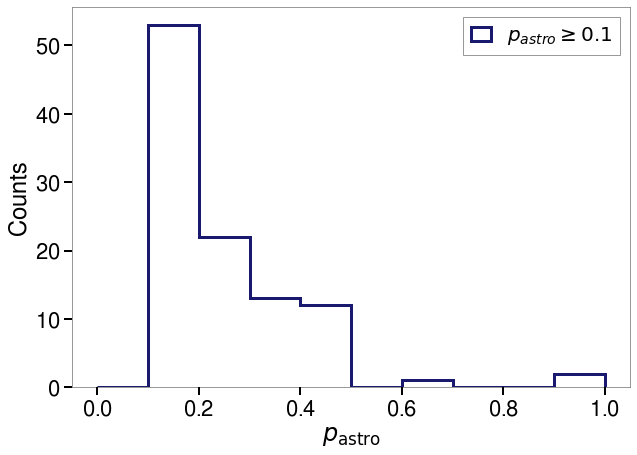}
    \caption{(Left) Distribution of $p_\mathrm{astro}$ values of GWs from GWTC-2.1 and GWTC-3, which are not LVK's confidently detected events. It includes both CBCs and GW burst-like events detected during O3. The last bin, highlighted in pale blue with hatching, includes events with $p_\mathrm{astro} \geq$ 0.1. There are 103 GWs, all selected for the archival neutrino counterpart search with sub-TeV neutrinos.
    (Right) Detailed $p_\mathrm{astro}$ distribution of the 103 shortlisted sub-threshold GWs.
    %within the range 0.1 $ \leq p_\mathrm{astro} \leq$ 0.5.
    Most of them have $p_\mathrm{astro} <$ 0.5, and 2 yr$^{-1} \leq $ FAR $<$ 2 day$^{-1}$. However, as shown in the plot, for three shortlisted GWs, $p_\mathrm{astro}$ exceeds 0.5. They also have FAR $<$ 2 yr$^{-1}$. All three were detected only by the cWB search pipeline and were not claimed as confidently detected events by LVK.
    }
    \label{fig:pastro}
\end{figure}

As part of this analysis, we exclude the GW alerts detected with high significance, based on their FAR values and detection times. For the remaining candidates, their $p_\mathrm{astro}$ distribution is shown in the left plot of Fig.~\ref{fig:pastro}. More than 90\% of these sub-threshold GWs were found to have $p_\mathrm{astro} \simeq 0$, indicating a high likelihood of being accidental triggers caused by terrestrial noise. Hence, we further refine our selection to events with $p_\mathrm{astro} \geq 0.1$. This yields 103 shortlisted GW candidates. Their $p_{\rm astro}$ distribution is shown in the right plot of Fig.~\ref{fig:pastro}. Three of them have $p_{\rm astro} >$ 0.5, all detected by the cWB-allSky search pipeline.
%preventing them from being labelled as confidently detected GW signals 
They have higher $p_{\rm astro}$ and lower FAR values compared to GW signals commonly classified as sub-threshold detections. But they were not classified as confident detections because they were not associated with any confident CBC signal.
%they were not associated with any confident CBC detections from O3. Hence, they were not labelled as confident detections.
We nevertheless consider them for our follow-up analysis due to their strong signal characteristics.

\begin{figure}[htb!]
    \centering
    \includegraphics[width=0.8\linewidth]{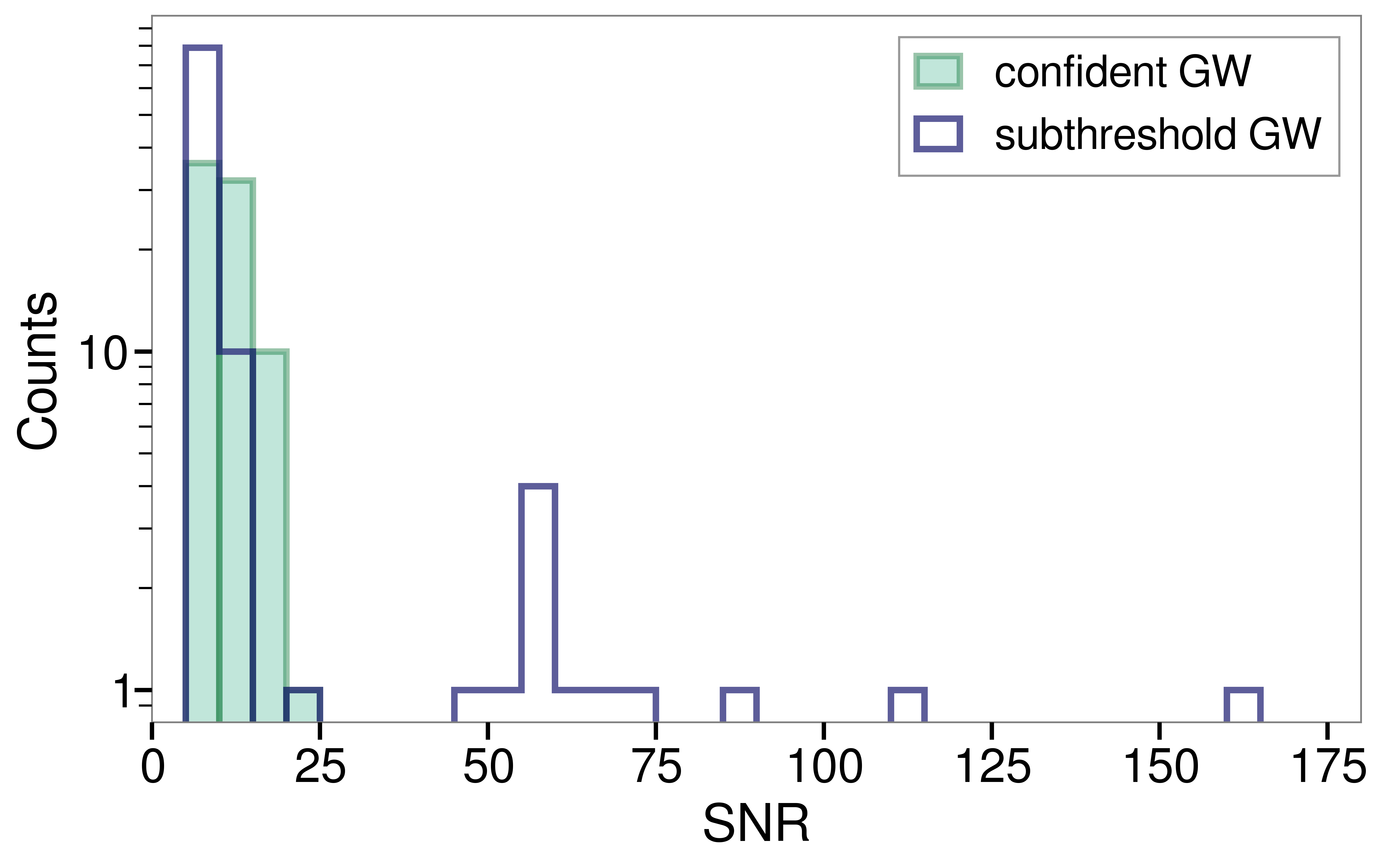}
    \caption{The SNR distribution of the 103 shortlisted sub-threshold GW events from O3. A comparison is shown with the SNR distribution of the 79 confidently detected GWs during O3.}
    \label{fig:snr}
\end{figure}

For the 103 shortlisted sub-threshold GW events, the signal-to-noise ratios (SNR) of CBCs are comparable to those of confidently detected CBCs from O3, as shown in Fig. \ref{fig:snr}. Overall, this indicates that the selected candidates are not intrinsically weak signals. In particular, it is prominent in Fig. \ref{fig:snr} that a subset exhibits notably high SNRs, exceeding 40. All of them are among the shortlisted cWBs. %The cWB SNR is a coherent network statistic and is not directly comparable to the matched-filter SNR of the CBC pipelines.
%Rather, their classification as low-significance detection is likely due to instrumental limitations, such as reduced detector sensitivity at the time of observation.
%Rather, their classification as low-significance detections is likely due to their signal morphology deviating from typical CBC waveform models.
%This results in a higher FAR in template-based searches.
%Such mismatches can arise if the signals originate from unconventional sources, for example.
%or are affected by instrumental artefacts, for example. 
%It is also evident that there is no correlation between event candidates having high SNR and those having high pastro. SNR measures directly come from the matched filtering techniques. In contrast, the pastro values are the Bayesian posteriors influenced by the signal model assumptions preferred by the modelled or minimally modelled search pipelines.
%For the 103 shortlisted sub-threshold GW events, the signal-to-noise ratios (SNRs) are comparable to those of confidently detected GWs from O3, as shown in Fig.~\ref{fig:snr}. This suggests that the selected candidates are not necessarily intrinsically weak signals. 
Their classification as low-significance detections may reflect their signal morphology deviating from typical CBC waveform models referred to in standard template-based and minimally-modelled searches. Such mismatches can occur, for example, for eccentric binary black hole mergers formed dynamically in dense stellar environments or AGN disks \citep{EccentricBBH_2019, EccentricBBH_O3}.
%where standard template-based searches are known to be less efficient

Additionally, it is important to identify the source type associated with each event. For the 103 shortlisted events, 13 are cWB detections. Among the 90 CBCs, there are 16 NSBH-like candidates. The rest are BBH-like events. The source classification is shown in Tab.~\ref{tab:source_classes}.

\begin{comment}
    \begin{figure}[htb!]
    \centering
    \includegraphics[width=0.8\linewidth]{plots/O3_subth_sources_paper.png}
    \caption{Different types of source classes in our GW shortlist. Most of them are BBH merger-like events. There are 16 NSBHs and no BNS-merger-like candidates. The remaining 13 are burst-like GW (cWB) candidates.}
    \label{fig:GW_source_types}
\end{figure}
\end{comment}

\begin{table}[h]
\centering
\caption{Source classification of the shortlisted 103 sub-threshold GW}
\label{tab:source_classes}
\begin{tabular}{lc}
\hline
Source class & Number of events \\
\hline
Binary Black Hole (BBH) & 74 \\
Neutron Star-Black Hole (NSBH) & 16 \\
GW burst-like events (cWB) & 13 \\
\hline
\end{tabular}
\end{table}

\section{Analysis Method}
\label{sec:UML}

To search for the neutrino counterparts to the individual shortlisted GWs, we employ an unbinned maximum likelihood (UML) analysis method. For each GW candidate, we define a search time window of 
$\pm$500\,s around the GW detection time, motivated by the timescales of neutrino emission expected from CBC, in analogy with short gamma-ray bursts \citep{CBC_GRB_timeWindow}. A box-shaped temporal profile is adopted, assigning equal probability to signal neutrino arrival within the search window and keeping it model-independent.
%Given the low event rate of the dataset, we expect only $\approx$4-5 events within the 1000\,s window.
We add a Poisson prefactor to the product term in the likelihood to account for the transient nature of the source,
%fluctuations in this small number of observed events,
making the method sensitive to a neutrino excess above background expectation in a short timescale. Assuming $N$ to be the total number of neutrino events observed within this time window,
we define the following likelihood function:
%$\mathcal{L}$ ($n_\mathrm{s}( \gamma$)), which is given as
\begin{equation}
\label{likelihood}
    \mathcal{L} (n_\mathrm{s} (\gamma)) = \frac{(n_\mathrm{s} + n_\mathrm{b})^N}{N!} \mathrm e^{-(n_\mathrm{s} + n_\mathrm{b})} \prod_{i = 1}^{N} \Big ( \frac{n_\mathrm{s} S_\mathrm{i}}{n_\mathrm{s} + n_\mathrm{b}} + \frac{n_\mathrm{b} B_\mathrm{i}}{n_\mathrm{s} + n_\mathrm{b}} \Big ).
\end{equation}

%Here, N is the total number of events observed \tm{within the search time window}.
The Poisson mean of the number of signal neutrinos is $n_\mathrm{s}$, and $n_\mathrm{b}$ is the expected number of background events.
%The index $i$ runs over each of the $N$ candidate neutrino events.
The signal and background probability density functions (PDF) for the $i^\mathrm{{th}}$ event are denoted as $S_\mathrm{i}$ and $B_\mathrm{i}$, respectively. Each consists of spatial and energy-dependent terms, as shown in Eq.~\ref{eq:Si_Bi}:

\begin{equation}
S_\mathrm{i}
%\equiv S(\mathbf{\Omega}i, E_i, \sigma_i \mid \mathbf{\Omega}_\mathrm{src}, \gamma)
= S^\mathrm{space}(\mathbf{x}_\mathrm{i}, \sigma_\mathrm{i} \mid \mathbf{x}_\mathrm{src}) \cdot S^\mathrm{energy}(E_\mathrm{i} \mid \gamma),
\\
B_i
%\equiv B(\delta_i, E_i)
= B^\mathrm{space}(\delta_\mathrm{i}) \cdot B^\mathrm{energy}(E_\mathrm{i} \mid \delta_\mathrm{i}).
\label{eq:Si_Bi}
\end{equation}

%are defined following the method explained in \cite{greco-gw, greco_Erratum}.
%Each of them consists of spatial and energy-dependent terms.
The signal PDF for the $i^\mathrm{{th}}$ event, $S_\mathrm{i}$, is constructed using event-level information, such as the reconstructed sky position ($\mathbf{x}_i$), energy ($E_i$), and angular uncertainty ($\sigma_i$). It also incorporates information about the source position ($\mathbf{x}_\mathrm{src}$) and depends on the spectral index $\gamma$, which characterises the assumed energy spectrum of the signal neutrinos.
%There is a temporal component as well, which represents the time window in which signal neutrino emission is expected from individual sources.

The background PDF, $B_\mathrm{i}$, for the $i^\mathrm{{th}}$ event is estimated from scrambled data and depends primarily on the event’s declination ($\delta_i$) and reconstructed energy ($E_i$).

\begin{comment}
    The signal PDF for the $i^\mathrm{{th}}$ event, $S_\mathrm{i}$, is constructed using the reconstructed sky position ($\mathbf{x_i}$), energy ($E_i$), and angular uncertainty ($\sigma_i$) of the event. It also incorporates information about the source position ($\mathbf{x_s}$) and depends on the spectral index $\gamma$, which characterises the assumed energy spectrum of the signal neutrinos.
The spatial component is modelled using the 3D von Mises-Fisher distribution, given as,
\begin{equation}
    S_{i, \rm space} (\mathbf{x_i, x_s}, \sigma) = \frac{\kappa_i}{4 \pi~\rm sinh~\kappa_i} \rm exp (\kappa_i~cos(|\mathbf{x_i - x_s|})) 
    \label{eq:spatial_signal}
\end{equation}

In Eq.~\ref{eq:spatial_signal}, $\kappa_i = 1/\sigma_i$.
\end{comment}

Using this likelihood function, we define a test statistic (TS) for a source at the $k^\mathrm{th}$ pixel by maximising the likelihood with respect to the free parameters $n_\mathrm{s}$ and $\gamma$.
%we define a test statistic (TS) evaluated at each sky pixel `k':

\begin{equation}
\label{eq:TS}
\mathrm{TS_k} =\max_{n_\mathrm{s},\,\gamma}
\Bigg[ 2~\mathrm{ln}  \Bigg( \frac{\mathcal{L}_\mathrm{k} (n_\mathrm{s}(\gamma))}{\mathcal{L}_\mathrm{k} (n_\mathrm{s} = 0)} \Bigg) + 2~\ln(\omega_k)\Bigg ].
\end{equation}

We use Hierarchical Equal Area isoLatitude Pixelization (HEALPix)~\citep{Healpix} to split the sky into equal-area pixels, and $\omega_k$ is the spatial prior associated with the $k^\mathrm{th}$ pixel. We evaluate it as the per-pixel probability from the individual GW skymap, normalised to the skymap's maximum-probability pixel.
%This ensures $0 < \omega_k \leq 1$ and the maximum-probability pixel has $\omega_k = 1$.
This ensures that the spatial penalty term, 2~ln($\omega_k$) = 0 at the maximum-probability pixel from the GW skymap.

For each neutrino event, whether scrambled or from the data, within the search time window, we scan the HEALPix pixels which lie within the 3$\sigma$ radius around each event's reconstructed direction.
%Thus, the set of scanned pixels comprises the sky positions considered as possible source locations in the likelihood analysis.
For each scanned pixel $k$, the test statistic $\mathrm{TS}_k$ is evaluated by maximising the likelihood over the signal parameters. The final TS is then obtained by taking the maximum of the per-pixel TS over all scanned pixels~\citep{greco-gw, greco_Erratum}.
\begin{equation}
\label{eq:TS_final}
\mathrm{TS} = \max_\mathrm{k} \mathrm{TS_k}.
\end{equation}
%We evaluate the TS at each scanned pixel, following Eq.~\ref{eq:TS}, by maximising the likelihood with respect to the free parameters $n_\mathrm{s}$ and $\gamma$ \citep{greco-gw, greco_Erratum}. The final TS reported then is the maximum TS over all scanned sky pixels.
%The TS is evaluated at each scanned pixel by maximising over $n_\mathrm{s}$ and $\gamma$, including the spatial penalty. The same sky-pixel scan and TS calculation are applied to the observed data and to each background pseudo-experiment.

%Here, $k$ is the index of each pixel in the sky. In Eq.~\ref{eq:TS}, $\omega_k$ is the spatial prior associated with that pixel.
%$\omega_{k}$ is a spatial prior term at that location. For each pixel, $\omega_\mathrm{k}$ is proportional to the probability of having a GW source location in that pixel. At each sky pixel, the likelihood is maximised with respect to the free parameters $n_\mathrm{s}$ and $\gamma$. The final TS reported for each GW event is the maximum TS over all sky pixels.

To test the hypotheses, the background TS distribution for each sub-threshold GW candidate is constructed by randomly sampling the arrival times of neutrino events detected within $\pm$ 5 days of the time the GW signal was recorded. This task is repeated 10,000 times, which are designated as individual `pseudo experiments'. The maximum TS from each background trial is identified by scanning the pixels around scrambled events, resulting in a background TS distribution. It is then compared with the TS values from an unscrambled neutrino dataset to search for correlated signal neutrino events, as discussed in Sec.~\ref{sec:results}. The same background distribution is also used to evaluate the sensitivity of the analysis by injecting signal-like neutrino events assuming an $E^{-2}$ spectrum within a 1000~s ($\pm$500 s) time window. The choice of an $E^{-2}$ spectrum is motivated by the expected spectral index of CRs at the source, which the signal neutrinos should mimic closely. For each signal injection, a source position is randomly drawn from the GW skymap. The injected signal at that position corresponds to a neutrino flux recorded by the detector. This step is repeated 1000 times for each signal strength, considering individual sub-threshold GWs. The 90\% sensitivity is defined as the time-integrated neutrino flux for which 90\% of the injected pseudo-experiments yield a TS exceeding the median of the background TS distribution for the given GW event.
%But before doing that, we tested the strength of our analysis by `injecting' signal-like neutrino events within a 1000~s ($\pm$ 500~s) time window, from a Monte-Carlo dataset with $\gamma = 2$ when compared to a background-like, scrambled dataset. We choose the 1000~s search time window based on predicted neutrino emission from gamma-ray bursts triggered by CBC (\cite{CBC_GRB_timeWindow}).
%At each flux level, the Passing Fraction (PF) is the fraction of pseudo-experiments in which, for neutrino injection, the TS exceeds the median of the TS distribution for background. From such signal injections, we calculate the 90\% sensitivity for each shortlisted sub-threshold GW candidate. It is defined as the flux for which the passing fraction is 90\%. We calculate this by repeatedly choosing a source position according to the GW skymap and injecting neutrino events with mean $n_\mathrm{s}$ from the selected position. These injections give a set of TS values for $n_\mathrm{s}$ which can be compared to the background TS distribution.
Details of the analysis framework for sub-threshold GW and their sensitivity results have been presented \citep{ICRC23_proceeding, ECRS24_TM}.

\begin{figure}[htb!]
    \centering
    \includegraphics[width=0.8\linewidth]{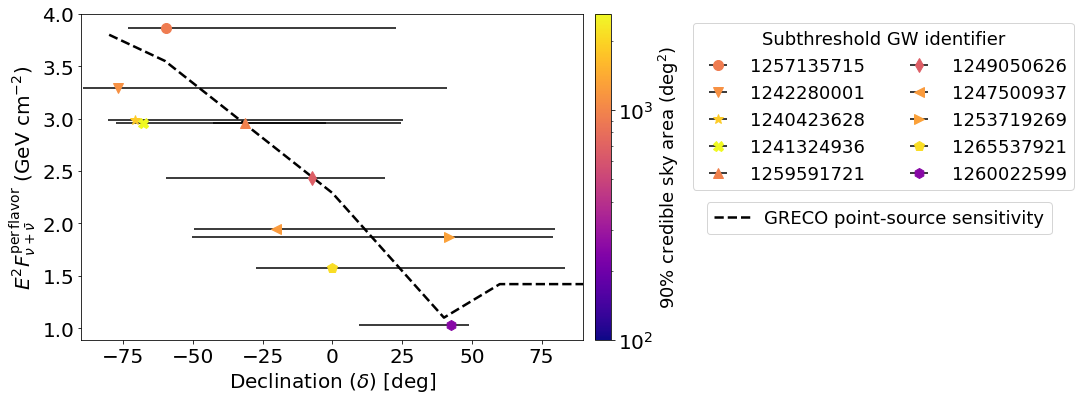}
    \caption{Declination dependence of per-flavour sensitivity flux calculated at 1 GeV for the sub-threshold CBC candidates identified by MBTA during O3. The declination of the pixel containing the most probable source location, as indicated by the GW skymap, is indicated by a marker. Marker colours encode the associated 90\% credible sky-localisation area for the respective GW candidates, identified by the GPS time of detection. The horizontal spread across the markers encompasses the declination span of the 90\% credible region in the respective GW skymap. Additionally, the overlaid dashed line shows the GRECO point source sensitivity at different declination values assuming a spectral index of 2, during a 1000~s time window.}
    \label{fig:GW_sens_dec}
\end{figure}

We also study the dependence of per-flavour sensitivity on declination for sub-threshold GW candidates. In Fig. \ref{fig:GW_sens_dec}, we show the sensitivities of the sub-threshold GWs identified by the MBTA pipeline, as these are some of the precisely localised sub-threshold GWs from O3. The y-axis represents the time-integrated energy-squared flux, where the per-flavour, neutrino and anti-neutrino summed neutrino flux is defined as,

\begin{equation}
    F_{\nu + \bar{\nu}}^{\mathrm{per-flavour}} = \frac{\mathrm{d}N}{\mathrm{d}E~\mathrm{d}A~\mathrm{d}t} \Delta t = \phi_0 \cdot \Big( \frac{E}{E_0} \Big)^{-\gamma}.
\end{equation}

It has the units of GeV~$^{-1}$~cm$^{-2}$. Here, $N$ is the number of neutrinos, $E$ is the energy, $A$ is the area, and $\Delta t$ = 1000~s. The flux normalisation, $\phi_0$ is at the reference energy, $E_0$ = 1~GeV. The sky localisation area derived from the 90\% credible region of the individual GW skymap is indicated by the colour of each marker, for which the sensitivity results are shown here. From this figure, we see that the declination dependence of sensitivity for poorly localised sources, such as the considered GW events, closely follows that of point sources. This indicates that the large localisation uncertainties of the GW events do not significantly degrade the sensitivity. The event-to-event sensitivity variation is driven primarily by the declination-dependent detector performance.
%The sensitivity scale is also very similar.
%It is primarily determined by detector performance as a function of declination rather than by the GW localisation area. Consequently, the event-to-event variation in sensitivity is dominated by source declination. As observed in previous GRECO analyses~\cite{greco-gw, greco_Erratum}, the sensitivity is better in the Northern Hemisphere than in the Southern Hemisphere. Furthermore, several sub-threshold GW candidates have 90\% localisation regions that span both hemispheres, reflecting their large sky-position uncertainties. Identifying spatially and temporally coincident neutrino counterparts would substantially improve the localisation of such GW candidates.
%This indicates that poor localisation of both the GW and the GRECO sample does not degrade sensitivity. The event-to-event variation in sensitivity is primarily due to source declination, determined by detector performance as a function of declination.
%rather than the GW localisation area.

%the general behaviour for the declination dependence of sensitivity for extended sources, like the considered GW events, follows that of the point-source sensitivity.
Consequently, sensitivity is better in the Northern Hemisphere than in the Southern Hemisphere.
%as identified in a previous analysis (\cite{greco-gw, greco_Erratum}).
Furthermore, it is notable that even the 90\% containment region for some sub-threshold GWs is relatively large, spanning both hemispheres. %The main reason for this is the worse localisation of GW sources by the GW detectors.
Identifying spatially and temporally coincident neutrino counterparts would significantly improve the localisation of such GW events.

\section{Results}
\label{sec:results}

To identify neutrinos spatially and temporally correlated to the 103 shortlisted GW events, we conduct a catalogue search considering a $\pm$500 s time window around the GW event detection time.
%The results of this analysis are summarised here.

\begin{figure}[htb!]
    \centering
    \includegraphics[width=0.8\linewidth]{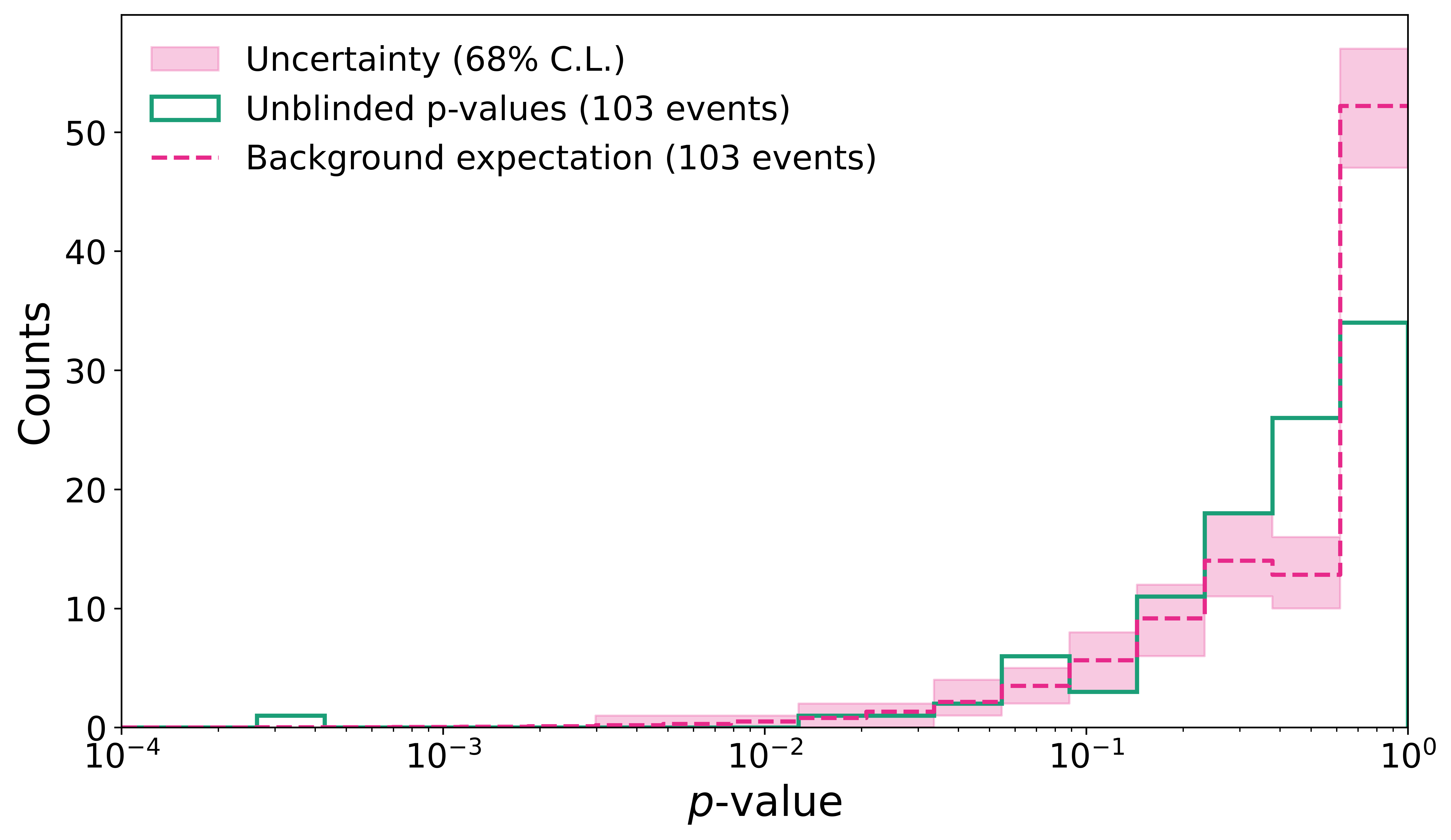}
    \caption{Distribution of pre-trial $p$-values for the 103 shortlisted GWs from O3. It is compared with the expected background distribution, obtained by randomly sampling TS values from the background TS distribution for each GW event and calculating the corresponding $p$-values. This background sampling was repeated 10,000 times to construct the expectation. The associated error band represent the 68\% error around the background expectation. The green histogram shows the observed $p$-value distribution from the UML method. The observed distribution is consistent with the null hypothesis, and no significant excess is observed.}
    \label{fig:pval_dist}
\end{figure}

Fig. \ref{fig:pval_dist} shows the distribution of pre-trial $ p$-values.
To assess its consistency with background expectations, we randomly sample TS values from the background TS distribution of each GW event and compute the corresponding $p$-values. We repeat this procedure 10,000 times for each GW event, and the background expectation at each $p$-value bin is marked with the magenta histogram in Fig.~\ref{fig:pval_dist}.
%To interpret whether this distribution is consistent with background expectations, a reference background distribution is established for comparison. It is obtained by randomly sampling TS values 10,000 times from the background TS distribution for each GW event and then computing the corresponding $p$-values.
%the mean of the accumulated $p$-values at each $p$-value bin provides a background expectation. It is marked with the magenta histogram in Fig.~\ref{fig:pval_dist}.
It is then compared with the observed $p$-value distribution for the 103 shortlisted GW events obtained with the UML method, which is overlaid as the green histogram.
%It is found to be consistent with the background expectation.
%the accumulated $p$-values are binned. The mean count in each p-value bin defines the expected background in that range.
%and the central 68\% interval representing the expected statistical fluctuations.
It is characterised by a pronounced pile-up at high $p$-values while bins near $p$ = 0 remain sparsely populated. This behaviour arises because of the extended nature of the GW skymaps and the poor angular resolution of GRECO events. Consequently, the likelihood frequently favours the background-only hypothesis. Also, within $\pm$500 s, we are in a low-background regime, where the number of events can only be a few integer values. This results in discretisation of the TS distribution, and the corresponding $p$-value distribution develops features, as visible in Fig. \ref{fig:pval_dist}, instead of being smooth and uniform.
%leading to distinct features in the background expectation.
%In such situations, the background TS distribution effectively reduces to a counting experiment, resulting in the corresponding $p$-value distribution developing features, instead of being smooth and uniform, as visible in Fig. \ref{fig:pval_dist}. This behaviour is particularly pronounced near $p$ = 1.0, where most pseudo-experiments yield low TS values.

\begin{figure}[htb!]
    \centering
    \includegraphics[width=0.49\textwidth]{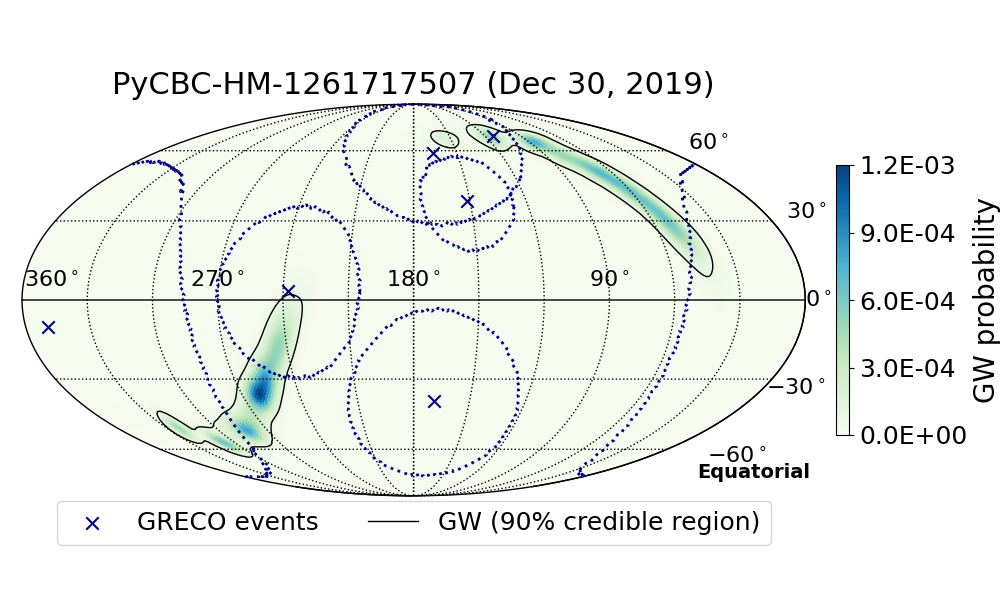}
    \includegraphics[width=0.49\textwidth]{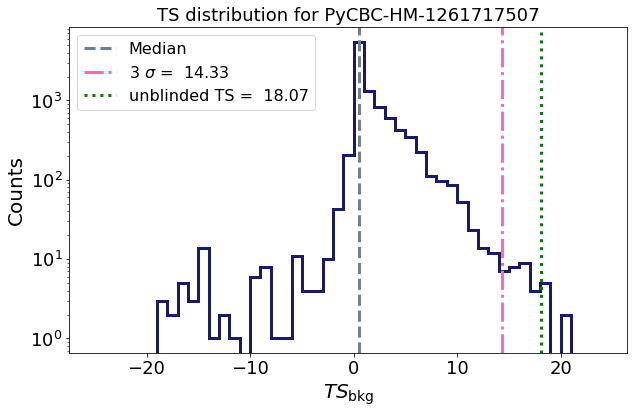}
    \caption{(Left) Skymap of the most significant sub-threshold GW from the shortlisted 103 events. The GPS time for the detection of the GW event is 1261717507, corresponding to December 30, 2019. The $p_{\rm astro}$ value corresponding to this detection is 0.16. The neutrinos detected within 1000 s of the sub-threshold GW detection time are overlaid on the GW skymap with individual crosses. The circles around each cross represent the 90\% angular uncertainty associated with each detected event. The probability of chance coincidence for this event is 0.04\% (4\% after trial correction). (Right) The observed TS value for PyCBC-HM-1261717507, overlaid on the corresponding background TS distribution. The median of the background distribution and 3$\sigma$ deviation from it are also indicated.}
    \label{fig:joint_GW_nu_skymap}
\end{figure}

%As the observed $p$-values are compared with the corresponding background expectation, no significant deviation from the null hypothesis is apparent.
%It follows the general features of the background expectation.
The observed distribution is overall consistent with the background, with the smallest pre-trial $p$-value $\sim 0.0004~(3.35\sigma)$, related to a sub-threshold BBH-like event. It shows the strongest correlation with the detected GRECO events within the 1000~s search time window. The GPS time of the event is 1261717507. It was detected by the PyCBC-Highmass pipeline on December 30, 2019, during O3b. However, after trial correction using the Šidák correction method \citep{sidak_correction}, the $p$-value drops to 0.04 (1.76$\sigma$). The joint skymap for the GW and GRECO events detected within 1000~s by IceCube is shown in Fig. \ref{fig:joint_GW_nu_skymap}. The right panel shows the observed TS value for this event, overlaid on the background TS distributions derived from scrambled trials. The background TS distribution has a negative tail, as many trials fall on pixels with high spatial penalty.
%Since the background scan is restricted to pixels within $3\sigma$ of the scrambled events, the maximum-probability GW pixel is not necessarily included. Thus, the maximum penalised TS can be negative.
The median and the $3\sigma$ deviation from the background are marked for reference. More example skymaps are shown in Fig.~\ref{fig:unblinded_skymaps_most_sig} and Fig.~\ref{fig:unblinded_skymaps_least_sig}.

\begin{figure}[htb!]
    \centering
    \includegraphics[width=0.8\linewidth]{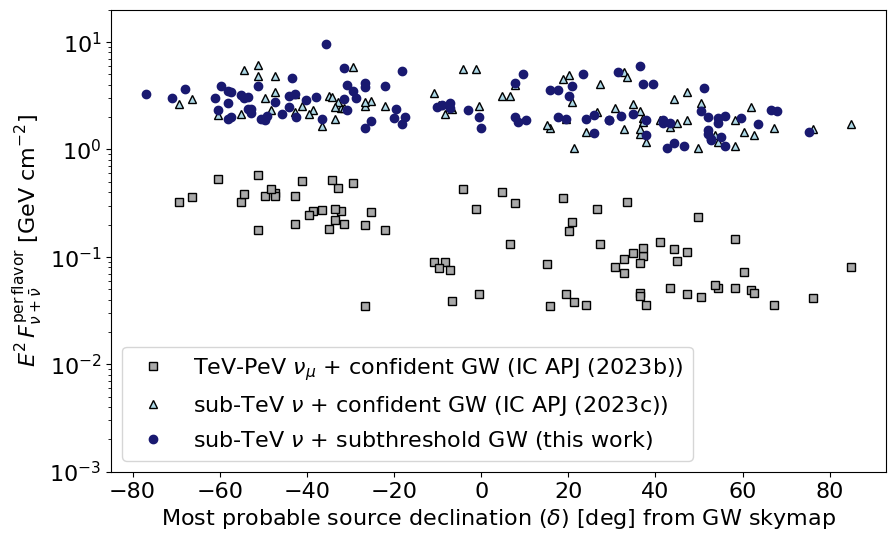}
    \caption{90\% C.L. flux upper limits for neutrino emission associated with the 103 shortlisted sub-threshold GW events. The calculation has been performed assuming a spectral index, $\gamma$ = 2 and a 1000~s integration time. Results are compared to previously published upper limits for confident GW events from O3, using both sub-TeV all-flavour (GRECO)~\citep{greco-gw, greco_Erratum} and TeV-PeV tracks (GFU)~\citep{GWfollowupO3} samples. Each flux value is plotted against the most probable source declination obtained from the GW skymap.}
    \label{fig:flux_UL}
\end{figure}

As none of the shortlisted events has more than 3$\sigma$ significance after trial correction, we report flux upper limits (90\% C.L.) for all of them. These limits apply to neutrino emission with an $E^{-2}$ spectrum during a 1000~s time window centred on the GW detection time.
%These limits were calculated assuming an $E^{-2}$ neutrino spectrum and a fixed 1000~s integration time centred on each GW detection time.
The flux limits are plotted with respect to the best-fit declination obtained solely from the corresponding GW skymaps in Fig. \ref{fig:flux_UL}. For context, we also include previously published limits for 79 confident GW events from O3, obtained using both the high-energy track (GFU) datasets and GRECO \citep{GWfollowupO3, greco-gw, greco_Erratum}. The comparison shows that the limits within the sub-TeV neutrino energy range with GRECO for sub-threshold GWs are of a similar order of magnitude to those for confident events. The limits obtained with GFU are more constraining, as expected, due to improved angular resolution for high-energy tracks.
This enables more effective suppression of background within a smaller search region. This enhances signal-to-background separation and yields better constraints. The limits for all 103 sub-threshold GWs and the corresponding $p_{\rm pre}$ values are summarised in Tab.~\ref{tab:unblinding_results}.

\startlongtable
\begin{deluxetable*}{lllcccc}
\tablecaption{Summary of unblinding results for the shortlisted 103 GW events. The events are arranged in ascending order of the corresponding pre-trial $p$-value ($p_\mathrm{pre}$). For each event, the GW event details, time of occurrence, area of 90\% credible region from the GW skymap, pre-trial $p$-value, and time-integrated flux upper limit $E^2F_{\nu+\bar{\nu}}^{\rm per-flavour}$ (for $\Delta t = 1000$~s) are listed.}
\tablehead{
\colhead{GW ID} & \colhead{Time (UTC)} & \colhead{Catalogue} & \colhead{Event-type} & \colhead{90\% credible area} & \colhead{$p_{\rm pre}$} & \colhead{Flux upper limit} \\
\nocolhead{} & \nocolhead{} & \nocolhead{} & \nocolhead{} &
\colhead{(deg$^2$)} & \nocolhead{} & \colhead{per-flavour} \\ \nocolhead{} & \nocolhead{} & \nocolhead{} & \nocolhead{} &
\colhead{} & \nocolhead{} & \colhead{$\mathrm{GeV\,cm^{-2}}$, for $\gamma = 2$}
}
\startdata
PyCBC-HM-1261717507 & 30.12.2019 & GWTC-3 & BBH & 2398.57 & 0.0004 & 9.65
\\
PyCBC-HM-1244202953 & 10.06.2019 & GWTC-2.1 & BBH & 2065.18 & 0.014 & 6.05
\\
PyCBC-HM-1265364040 & 10.02.2020 & GWTC-3 & BBH & 4109.77 & 0.033 & 5.22
\\
cWB-1259959714 & 09.12.2019 & GWTC-3 & cWB & 7019.54 & 0.039 & 5.05
\\
PyCBC-HM-1267132237 & 01.03.2020 & GWTC-3 & BBH & 2902.42 & 0.052 & 5.00
\\
cWB-1260301231 & 13.12.2019 & GWTC-3 & cWB & 2899.00 & 0.057 & 5.73
\\
GstLAL-1246849694 & 11.07.2019 & GWTC-2.1 & BBH & 3389.41 & 0.069 & 3.86
\\
GstLAL-1268338066 & 15.03.2020 & GWTC-3 & BBH & 10601.70 & 0.078 & 4.12
\\
PyCBC-HM-1238442602 & 04.04.2019 & GWTC-2.1 & BBH & 5624.87 & 0.078 & 5.32
\\
PyCBC-HM-1238423132 & 04.04.2019 & GWTC-2.1 & BBH & 1993.92 & 0.085 & 4.63
\\
PyCBC-HM-1244278698 & 11.06.2019 & GWTC-2.1 & BBH & 2015.56 & 0.086 & 4.04
\\
MBTA-1257135715 & 07.11.2019 & GWTC-3 & BBH & 917.85 & 0.1 & 3.86
\\
cWB-1248942961\tablenotemark{$\dagger$} & 04.08.2019 & GWTC-3 & cWB & 1753.42
& 0.13 & 3.99
\\
PyCBC-HM-1253402832 & 24.09.2019 & GWTC-2.1 & BBH & 5067.83 & 0.135 & 4.07
\\
cWB-1269257119 & 26.03.2020 & GWTC-3 & cWB & 2670.79 & 0.17 & 3.60
\\
PyCBC-HM-1243748255 & 05.06.2019 & GWTC-2.1 & BBH & 1637.25 & 0.17 & 3.54
\\
PyCBC-1252672417 & 16.09.2019 & GWTC-2.1 & BBH & 1556.37 & 0.17 & 3.72
\\
PyCBC-HM-1264333383 & 29.01.2020 & GWTC-3 & BBH & 5331.33 & 0.17 & 3.57
\\
cWB-1266548742 & 24.02.2020 & GWTC-3 & cWB & 2433.12 & 0.18 & 4.12
\\
GstLAL-1252939489 & 19.09.2019 & GWTC-2.1 & BBH & 1998.65 & 0.196 & 3.04
\\
PyCBC-1262657374 & 10.01.2020 & GWTC-3 & NSBH & 1622.90 & 0.21 & 3.88
\\
PyCBC-1263215647 & 16.01.2020 & GWTC-3 & NSBH & 3412.30 & 0.21 & 3.86
\\
cWB-1266055539 & 18.02.2020 & GWTC-3 & cWB & 2258.24 & 0.21 & 3.83
\\
GstLAL-1266028677 & 18.02.2020 & GWTC-3 & BBH & 3227.85 & 0.22 & 2.99
\\
PyCBC-1257260006 & 08.11.2019 & GWTC-3 & NSBH & 3340.62 & 0.2231 & 3.15
\\
PyCBC-HM-1243220837 & 30.05.2019 & GWTC-2.1 & BBH & 2840.54 & 0.2488 & 2.28
\\
PyCBC-HM-1239632437 & 18.04.2019 & GWTC-2.1 & BBH & 3147.11 & 0.249 & 1.74
\\
PyCBC-1251787804 & 06.09.2019 & GWTC-2.1 & NSBH & 2838.03 & 0.254 & 3.07
\\
PyCBC-1246272532 & 04.07.2019 & GWTC-2.1 & NSBH & 1671.11 & 0.26 & 3.54
\\
PyCBC-HM-1264569708 & 01.02.2020 & GWTC-3 & BBH & 1649.13 & 0.27 & 1.88
\\
cWB-1264222462 & 28.01.2020 & GWTC-3 & cWB & 4001.48 & 0.27 & 3.40
\\
GstLAL-1251926900 & 07.09.2019 & GWTC-2.1 & BBH & 2950.88 & 0.27 & 2.72
\\
MBTA-1241324936 & 08.05.2019 & GWTC-2.1 & BBH & 2634.49 & 0.28 & 3.62
\\
PyCBC-HM-1238821531 & 09.04.2019 & GWTC-2.1 & BBH & 7888.20 & 0.28 & 2.57
\\
GstLAL-1249479778 & 10.08.2019 & GWTC-2.1 & BBH & 2410.70& 0.2867 & 3.20
\\
PyCBC-1267485300 & 05.03.2020 & GWTC-3 & BBH & 1555.96 & 0.3016 & 2.07
\\
GstLAL-1266330503 & 21.02.2020 & GWTC-3 & BBH & 0.07 & 0.31 & 3.30
\\
PyCBC-HM-1244848599 & 17.06.2019 & GWTC-2.1 & BBH & 2228.40 & 0.34 & 2.46
\\
cWB-1250939932 & 27.08.2019 & GWTC-3 & cWB & 2348.70 & 0.34 & 2.12
\\
PyCBC-HM-1261149025 & 23.12.2019 & GWTC-3 & BBH & 6738.64 & 0.35 & 2.09
\\
MBTA-1259591721 & 05.12.2019 & GWTC-3 & BBH & 963.26 & 0.35 & 2.96
\\
PyCBC-HM-1250026325 & 16.08.2019 & GWTC-2.1 & BBH & 153.82 & 0.36 & 2.91
\\
PyCBC-1262924523 & 13.01.2020 & GWTC-3 & NSBH & 1725.21 & 0.37 & 1.99
\\
PyCBC-1262321982 & 06.01.2020 & GWTC-3 & NSBH & 2387.08 & 0.38 & 1.35
\\
MBTA-1240423628 & 27.04.2019 & GWTC-2.1 & BBH & 1847.27 & 0.39 & 2.99
\\
GstLAL-1241108686 & 05.05.2019 & GWTC-2.1 & BBH & 1017.62 & 0.40 & 2.75
\\
PyCBC-HM-1252645611 & 16.09.2019 & GWTC-2.1 & BBH & 17389.33 & 0.4277 & 3.06
\\
PyCBC-HM-1245839180 & 29.06.2019 & GWTC-2.1 & BBH & 4580.60 & 0.4292 & 2.00
\\
PyCBC-HM-1251890211 & 07.09.2019 & GWTC-2.1 & BBH & 6430.61 & 0.433 & 2.40
\\
PyCBC-1265685461 & 14.02.2020 & GWTC-3 & BBH & 2322.41 & 0.435 & 1.07
\\
PyCBC-HM-1265754804 & 14.02.2020 & GWTC-3 & BBH & 2913.04 & 0.447 & 1.32
\\
PyCBC-HM-1261197166 & 24.12.2019 & GWTC-3 & BBH & 1541.33 & 0.461 & 2.04
\\
PyCBC-1261058247 & 22.12.2019 & GWTC-3 & NSBH & 2905.13 & 0.4651 & 1.84
\\
cWB-1265755544\tablenotemark{$\dagger$} & 14.02.2020 & GWTC-3 & cWB &  1373.90 & 0.4709 & 2.37
\\
PyCBC-1249416363 & 09.08.2019 & GWTC-2.1 & NSBH & 3407.22 & 0.4876 & 2.18
\\
PyCBC-HM-1258872800 & 27.11.2019 & GWTC-3 & BBH & 2020.03 & 0.4881 & 1.94
\\
cWB-1253922430\tablenotemark{$\dagger$} & 30.09.2019 & GWTC-3 & cWB & 5196.76 & 0.4963 & 2.71
\\
PyCBC-1258003304 & 17.11.2019 & GWTC-3 & BBH & 3091.40 & 0.5065 & 2.05
\\
PyCBC-HM-1242314274 & 19.05.2019 & GWTC-2.1 & BBH & 5702.96 & 0.5087 & 1.79
\\
PyCBC-1264663254 & 02.02.2020 & GWTC-3 & NSBH & 2223.83 & 0.5105 & 1.87
\\
GstLAL-1266480585 & 23.02.2020 & GWTC-3 & BBH & 3653.02 & 0.5114 & 1.40
\\ 
cWB-1253293503 & 23.09.2019 & GWTC-2.1 & cWB & 2562.10 & 0.533 & 1.75
\\
PyCBC-1238582124 & 06.04.2019 & GWTC-2.1 & NSBH & 1112.37 & 0.5385 & 1.77
\\
PyCBC-HM-1264246793 & 28.01.2020 & GWTC-3 & BBH & 3272.73 & 0.5473 & 1.77
\\
PyCBC-HM-1261146225 & 23.12.2019 & GWTC-3 & BBH & 1744.93 & 0.5676 & 1.90
\\
PyCBC-1262314840 & 06.01.2020 & GWTC-3 & NSBH & 1907.22 & 0.5678 & 1.94
\\
GstLAL-1266516436 & 23.02.2020 & GWTC-3 & BBH & 2543.65 & 0.5715 & 1.26
\\
PyCBC-HM-1267131516 & 01.03.2020 & GWTC-3 & BBH & 5875.18 & 0.586 & 2.29
\\
PyCBC-1261656171 & 29.12.2019 & GWTC-3 & BBH & 1776.02 & 0.615 & 2.00
\\
PyCBC-HM-1263748131 & 22.01.2020 & GWTC-3 & BBH & 4255.23 & 0.6564 & 1.90
\\
cWB-1264014232 & 25.01.2020 & GWTC-3 & cWB & 4666.11 & 0.6918 & 2.34
\\
PyCBC-HM-1243985990 & 07.06.2019 & GWTC-2.1 & BBH & 2102.97 & 0.6995 & 1.93
\\
GstLAL-1240164426 & 24.04.2019 & GWTC-2.1 & BBH & 24219.44 & 0.7215 & 2.32
\\
PyCBC-HM-1261273562 & 25.12.2019 & GWTC-3 & BBH & 2627.54 & 0.8399 & 1.96
\\
PyCBC-1262800092 & 11.01.2020 & GWTC-3 & BBH & 1723.39 & 0.8577 & 1.41
\\
MBTA-1247500937 & 18.07.2019 & GWTC-2.1 & BBH & 1228.18 & 0.8711 & 1.95
\\
PyCBC-1268438406 & 16.03.2020 & GWTC-3 & NSBH & 1710.42 & 0.8825 & 1.73
\\
PyCBC-1244375744 & 12.06.2019 & GWTC-2.1 & NSBH & 1472.60 & 0.8884 & 3.04
\\
PyCBC-HM-1250205962 & 18.08.2019 & GWTC-2.1 & BBH & 2886.35 & 0.898 & 2.36
\\
PyCBC-HM-1241639735 & 11.05.2019 & GWTC-2.1 & BBH & 2447.76 & 0.90 & 1.60
\\
MBTA-1249050626 & 05.08.2019 & GWTC-2.1 & BBH & 663.45 & 0.90 & 2.43
\\
PyCBC-1243167760 & 29.05.2019 & GWTC-2.1 & BBH & 804.45 & 0.90 & 1.22
\\
PyCBC-HM-1268250455 & 14.03.2020 & GWTC-3 & BBH & 2316.24 & 0.91 & 1.46
\\
GstLAL-1261658694 & 29.12.2019 & GWTC-3 & BBH & 1907.05 & 0.91 & 3.12
\\
PyCBC-HM-1260305408 & 13.12.2019 & GWTC-3 & BBH & 1689.48 & 0.91 & 2.00
\\
PyCBC-HM-1258674399 & 24.11.2019 & GWTC-3 & BBH & 1624.47 & 0.92 & 1.52
\\
PyCBC-HM-1239092652 & 12.04.2019 & GWTC-2.1 & BBH & 1294.78 & 0.92 & 1.14
\\
PyCBC-1260407000 & 15.12.2019 & GWTC-3 & NSBH & 1798.23 & 0.93 & 1.88
\\
PyCBC-HM-1239227262 & 13.04.2019 & GWTC-2.1 & BBH & 15716.93 & 0.93 & 2.00
\\
cWB-1246523564 & 07.07.2019 & GWTC-3 & cWB & 6063.35 & 0.93 & 1.98
\\
MBTA-1265537921 & 12.02.2020 & GWTC-3 & BBH & 2143.03 & 0.94 & 1.57
\\
PyCBC-1245459820 & 25.06.2019 & GWTC-2.1 & BBH & 1016.98 & 0.94 & 2.00
\\
PyCBC-HM-1259379760 & 03.12.2019 & GWTC-3 & BBH & 1457.79 & 0.94 & 1.94
\\
PyCBC-1252064527 & 09.09.2019 & GWTC-2.1 & BBH & 4009.64 & 0.94 & 1.07
\\
PyCBC-HM-1249495257 & 10.08.2019 & GWTC-2.1 & BBH & 1966.12 & 0.94 & 1.94
\\
MBTA-1253719269 & 28.09.2019 & GWTC-2.1 & BBH & 1290.48 & 0.95 & 1.87
\\
GstLAL-1261725523 & 30.12.2019 & GWTC-3 & BBH & 5868.71 & 0.95 & 2.33
\\
PyCBC-HM-1244987689 & 19.06.2019 & GWTC-2.1 & BBH & 4746.57 & 0.96 & 2.02
\\
MBTA-1260022599 & 10.12.2019 & GWTC-3 & BBH & 250.35 & 0.96 & 1.03
\\
PyCBC-HM-1249097859 & 06.08.2019 & GWTC-2.1 & BBH & 4567.15 & 0.96 & 2.15
\\
MBTA-1242280001 & 19.05.2019 & GWTC-2.1 & BBH & 1132.20 & 0.97 & 3.29
\\
PyCBC-HM-1262878298 & 12.01.2020 & GWTC-3 & BBH & 2582.37 & 0.98 & 1.35
\\
PyCBC-1248215122 & 26.07.2019 & GWTC-2.1 & NSBH & 3126.51 & 0.99 & 2.46
\enddata
%\tablecomments{\textsuperscript{$\dagger$} cWB with $p_{\rm astro} > 0.5$.}
%\tablecomments{$^{\dagger}$ cWB with $p_{\rm astro} > 0.5$.}
%\tablerefs{$^{\dagger}$ cWB with $p_{\rm astro} > 0.5$.}
\label{tab:unblinding_results}
\end{deluxetable*}
{\footnotesize $^\dagger$ cWB with $p_{\rm astro} > 0.5$.}

\section{Summary and Discussion}
\label{sec:summary}

In this paper, we present results from the archival catalogue search following the UML method to identify $\mathcal{O}(10-100)$ GeV neutrino counterparts of 103 selected sub-threshold GW events. The search time window is 1000~s, centred on each GW detection time.
%The sub-threshold GWs, though lower in significance according to the GW detectors, are still promising source candidates for joint emission of GW and neutrino, which are not yet well identified.
%For sub-TeV neutrinos, the `GRECO Astronomy' dataset is used.
%which includes all-sky, all-flavour events detected with IceCube-DeepCore.
%The main motivation to perform this search was to understand the astrophysical origins of sub-threshold GWs better.
%As sub-threshold GWs represent valuable targets for multi-messenger searches, in this chapter, we present an archival UML follow-up search for sub-TeV neutrinos in temporal and spatial coincidence with sub-threshold GWs from O3.
No statistically significant neutrino counterpart was identified for any GW event. Although a BBH merger-like candidate had a neutrino coincidence with $p_\mathrm{pre} \sim$ 0.0004, it reduces to 1.76$\sigma$ after trial correction.
%The results disfavour the possibility of very strong $\mathcal{O}(10-100)$ GeV neutrino emission associated with the shortlisted sub-threshold GW events.
In the absence of a common source candidate of GW and neutrinos, neutrino flux upper limits were derived for all 103 GW events. The limits presented here complement those from the high-energy \citep{GWfollowupO3} and sub-TeV neutrino \citep{greco-gw, greco_Erratum} follow-ups, which were previously published. %Although the neutrino correlation is not significant after trial correction, this does not allow us to dismiss the credibility of the sub-threshold GWs as all of

Overall, the detection prospects are limited by the larger angular uncertainties of sub-TeV neutrinos, which degrade the signal-to-background separation and result in less constraining flux upper limits. These are currently insufficient to constrain or exclude source models. A major improvement in this aspect is foreseen with advanced neural network-based reconstruction methods, already under active development.
%Two complementary developments are expected to address this. Advanced neural network-based reconstruction methods, already under active development, will improve sub-TeV angular resolution.
The IceCube Upgrade \citep{IC_upgrade} will additionally provide higher exposure at lower energies, with improved localisation and energy determination. Together, these are expected to significantly enhance the sensitivity of GW follow-up searches in the sub-TeV regime~\citep{IC_Upgrade_sensitivity}.
%source identification capabilities and
%With improved reconstruction of sub-TeV neutrinos using advanced methods, such as neural networks, a major improvement in this aspect is foreseen. There are already multiple efforts in this direction. In addition, the IceCube Upgrade (\cite{IC_upgrade}) will provide higher exposure at lower energies and is expected to improve the detector's localisation and energy determination capabilities in the sub-TeV energy range. This is also likely to improve the capabilities of identifying astrophysical sources of sub-TeV neutrinos using IceCube data.

Looking ahead, the increasing sub-threshold GW detections from upcoming observation runs of LVK will provide us with the opportunity to probe more possible cases of joint emission.
%expand the sample of candidates for joint GW-neutrino searches.
Should third-generation detectors such as the Einstein Telescope become operational, improved source localisation and larger accessible detection volume will allow us to detect GW signals from distant astrophysical sources, with improved precision \citep{ET_BlueBook}. All these advancements will enhance the sensitivity of future catalogue searches for common astrophysical sources of GW and neutrinos. This analysis framework can also be readily extended to other IceCube data samples, including tracks-only or cascades-only datasets.
%With the increase in sub-threshold GW detection rates during the fourth observation run of LVK and the expected similarity of observations with upcoming third-generation GW detectors, we plan to conduct follow-up search campaigns, providing us with the opportunity to probe more possible cases of joint emission. With the help of the next-generation GW detectors, we also expect to detect GW signals coming from distant astrophysical sources, with improved precision and source localisation.  
%Better localisation of GW sources will also enhance the sensitivity of such a search.
With sub-threshold GW alerts now publicly available in real time during the fourth observation run of LVK, this work further opens promising avenues for future real-time multi-messenger campaigns, also with foreseen archival follow-up searches. This paper lays the groundwork for upcoming studies, the results of which will be addressed in future articles.

\begin{contribution}
The work presented in this article is based on significant contributions by Tista Mukherjee. It is based on data collected by the IceCube collaboration who designed, constructed and now operates the IceCube Neutrino Observatory. Data processing and calibration, Monte Carlo simulations of the detector and of theoretical models, and preparation and validation of the data samples were carried out through coordinated efforts across the collaboration. The collaboration approved the scientific results and was involved in preparing the manuscript.
\end{contribution}

\begin{acknowledgments}

The authors gratefully acknowledge the support from the following agencies and institutions:
USA {\textendash} U.S. National Science Foundation-Office of Polar Programs,
U.S. National Science Foundation-Physics Division,
U.S. National Science Foundation-EPSCoR,
U.S. National Science Foundation-Office of Advanced Cyberinfrastructure,
Wisconsin Alumni Research Foundation,
Center for High Throughput Computing (CHTC) at the University of Wisconsin{\textendash}Madison,
Open Science Grid (OSG),
Partnership to Advance Throughput Computing (PATh),
Advanced Cyberinfrastructure Coordination Ecosystem: Services {\&} Support (ACCESS),
Frontera and Ranch computing project at the Texas Advanced Computing Center,
U.S. Department of Energy-National Energy Research Scientific Computing Center,
Particle astrophysics research computing center at the University of Maryland,
Michigan State University,
Astroparticle physics computational facility at Marquette University,
NVIDIA Corporation,
and Google Cloud Platform;
Belgium {\textendash} Funds for Scientific Research (FRS-FNRS and FWO),
FWO Odysseus and Big Science programmes,
and Belgian Federal Science Policy Office (Belspo);
Germany {\textendash} Bundesministerium f{\"u}r Forschung, Technologie und Raumfahrt (BMFTR),
Deutsche Forschungsgemeinschaft (DFG),
Helmholtz Alliance for Astroparticle Physics (HAP),
Initiative and Networking Fund of the Helmholtz Association,
Deutsches Elektronen Synchrotron (DESY),
and High Performance Computing cluster of the RWTH Aachen;
Sweden {\textendash} Swedish Research Council,
Swedish Polar Research Secretariat,
National Academic Infrastructure for Supercomputing in Sweden (NAISS),
and Knut and Alice Wallenberg Foundation;
European Union {\textendash} EGI Advanced Computing for research;
Australia {\textendash} Australian Research Council;
Canada {\textendash} Natural Sciences and Engineering Research Council of Canada,
Calcul Qu{\'e}bec, Compute Ontario, Canada Foundation for Innovation, WestGrid, and Digital Research Alliance of Canada;
Denmark {\textendash} Villum Fonden, Carlsberg Foundation, and European Commission;
New Zealand {\textendash} Marsden Fund;
Japan {\textendash} Japan Society for Promotion of Science (JSPS), Ministry of Education, Culture, Sports, Science and Technology (MEXT), and Institute for Global Prominent Research (IGPR) of Chiba University;
Korea {\textendash} National Research Foundation of Korea (NRF);
Switzerland {\textendash} Swiss National Science Foundation (SNSF).
\end{acknowledgments}

%\facilities{HST(STIS), Swift(XRT and UVOT), AAVSO, CTIO:1.3m, CTIO:1.5m, CXO}

%\software{astropy \citep{2013A&A...558A..33A,2018AJ....156..123A,2022ApJ...935..167A},
%          Cloudy \citep{2013RMxAA..49..137F}, 
%          Source Extractor \citep{1996A&AS..117..393B}
%          }

\appendix

\section{Joint GW and neutrino skymaps}
We identified 13 GW events (see Tab.~\ref{tab:unblinding_results}) with pre-trial $p$-value $\leq 0.14 (\geq 1 \sigma)$. Fig.~\ref{fig:unblinded_skymaps_most_sig} shows their GW skymaps and the corresponding GRECO events detected within the 1000~s time windows for each GW event.

\begin{figure}[htb!]
    \centering
    \includegraphics[width=0.49\linewidth]{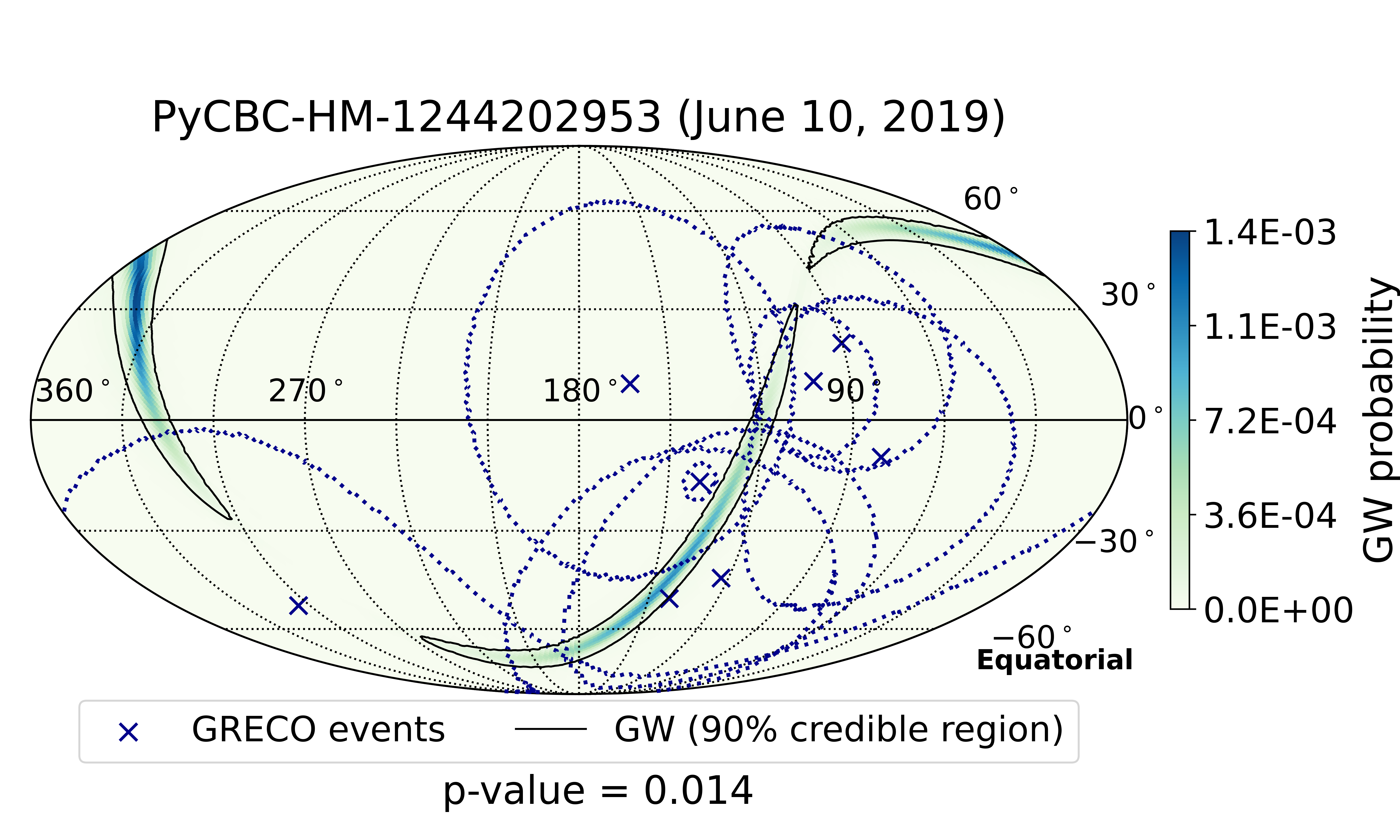}
    \includegraphics[width=0.49\linewidth]{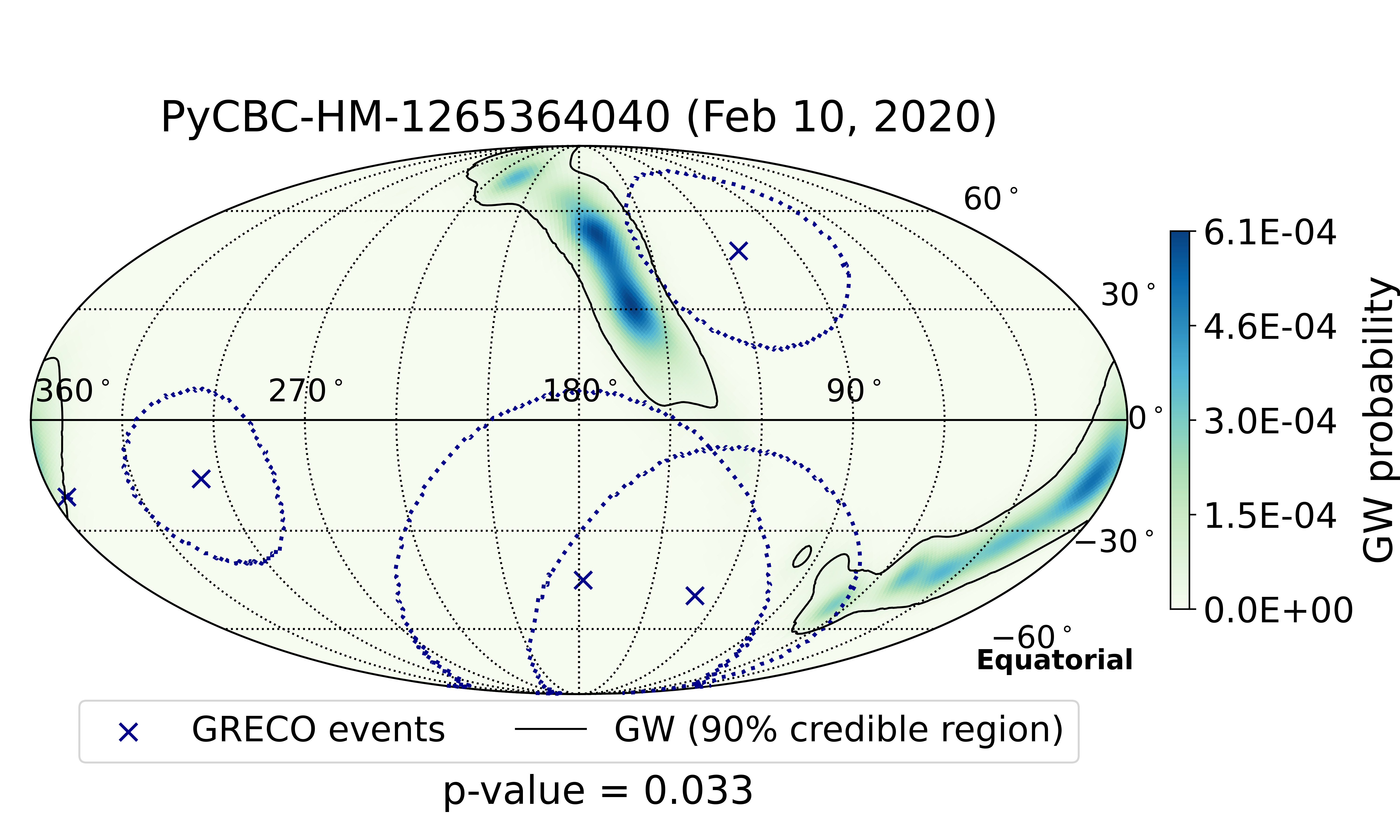}
    \includegraphics[width=0.49\linewidth]{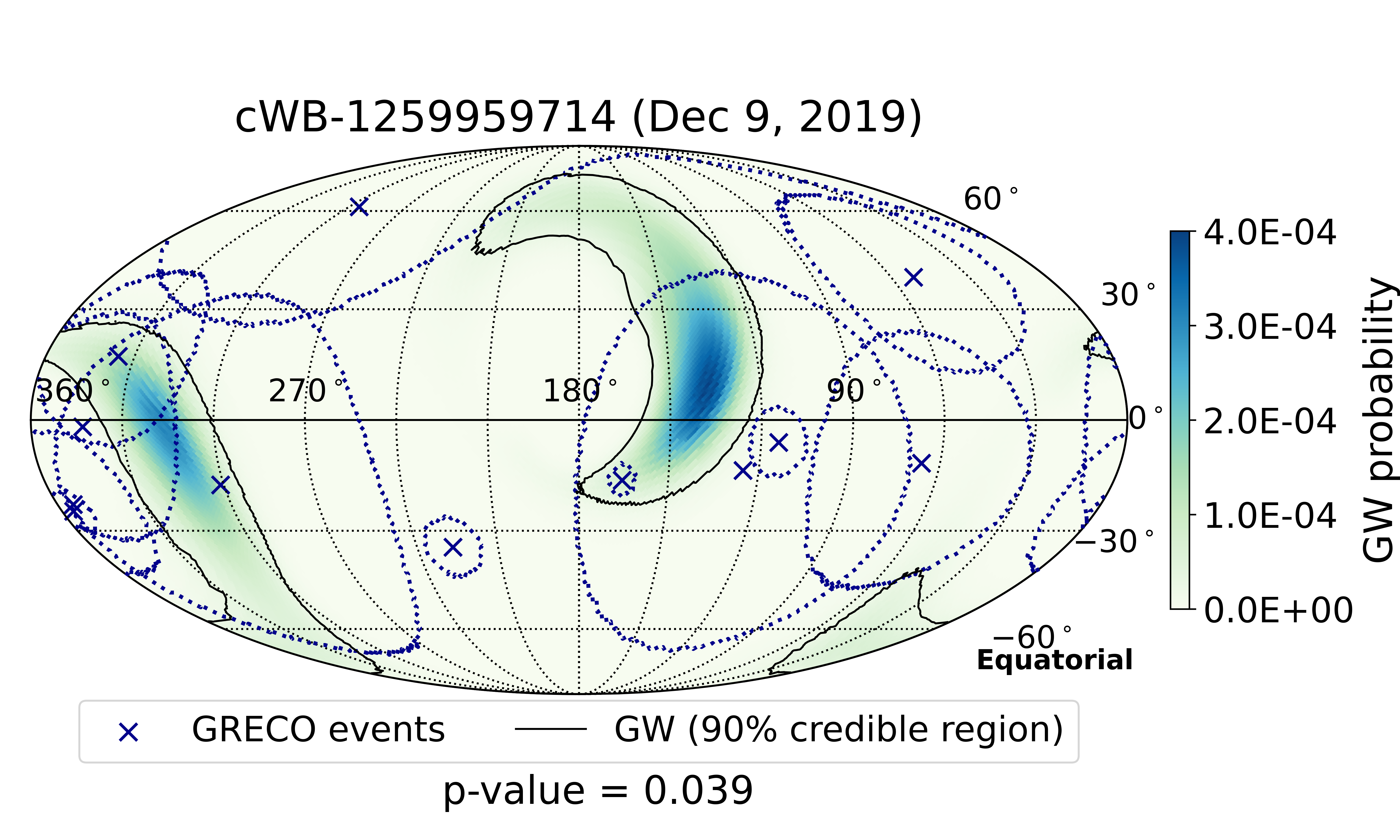}
    \includegraphics[width=0.49\linewidth]{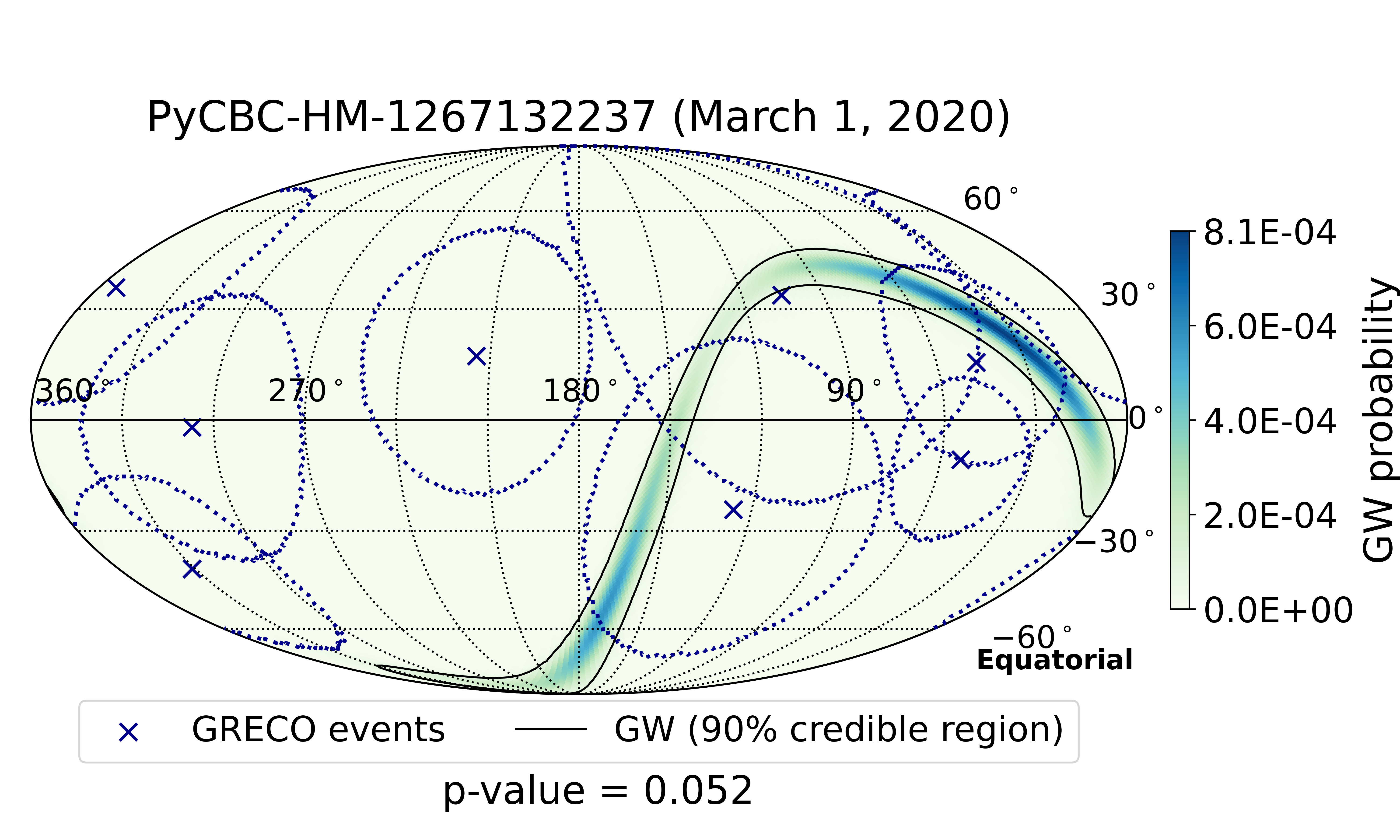}
    \includegraphics[width=0.49\linewidth]{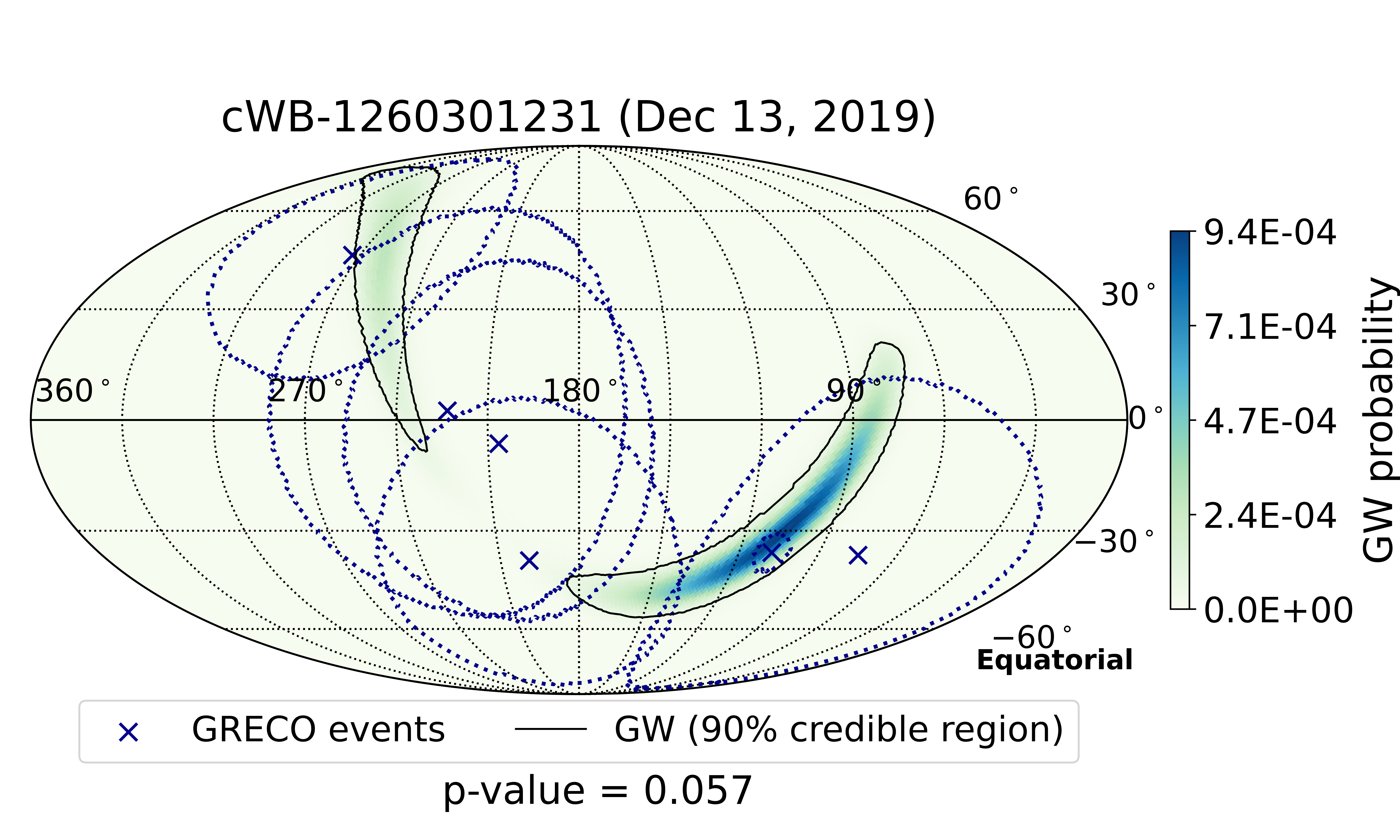}
    \includegraphics[width=0.49\linewidth]{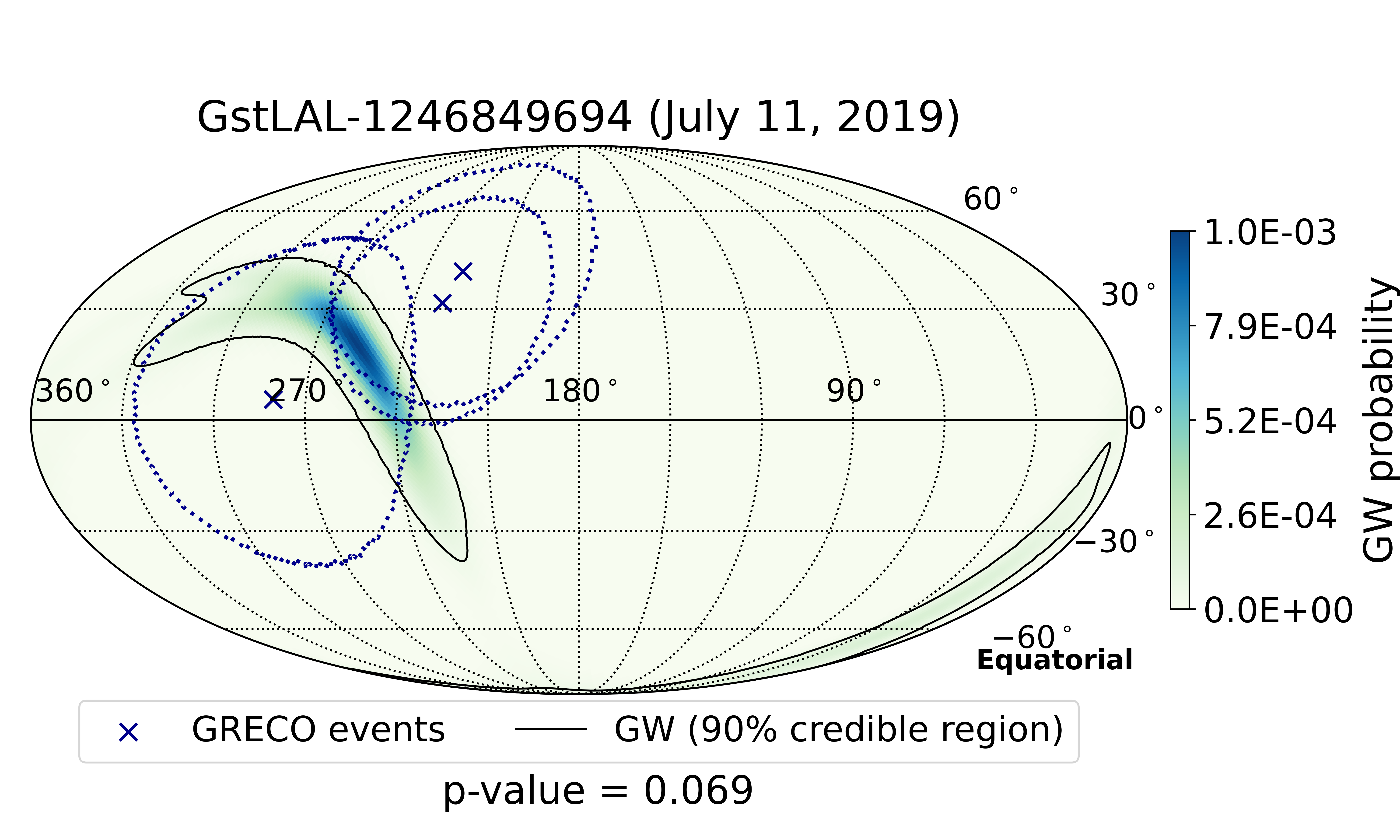}
    \includegraphics[width=0.49\linewidth]{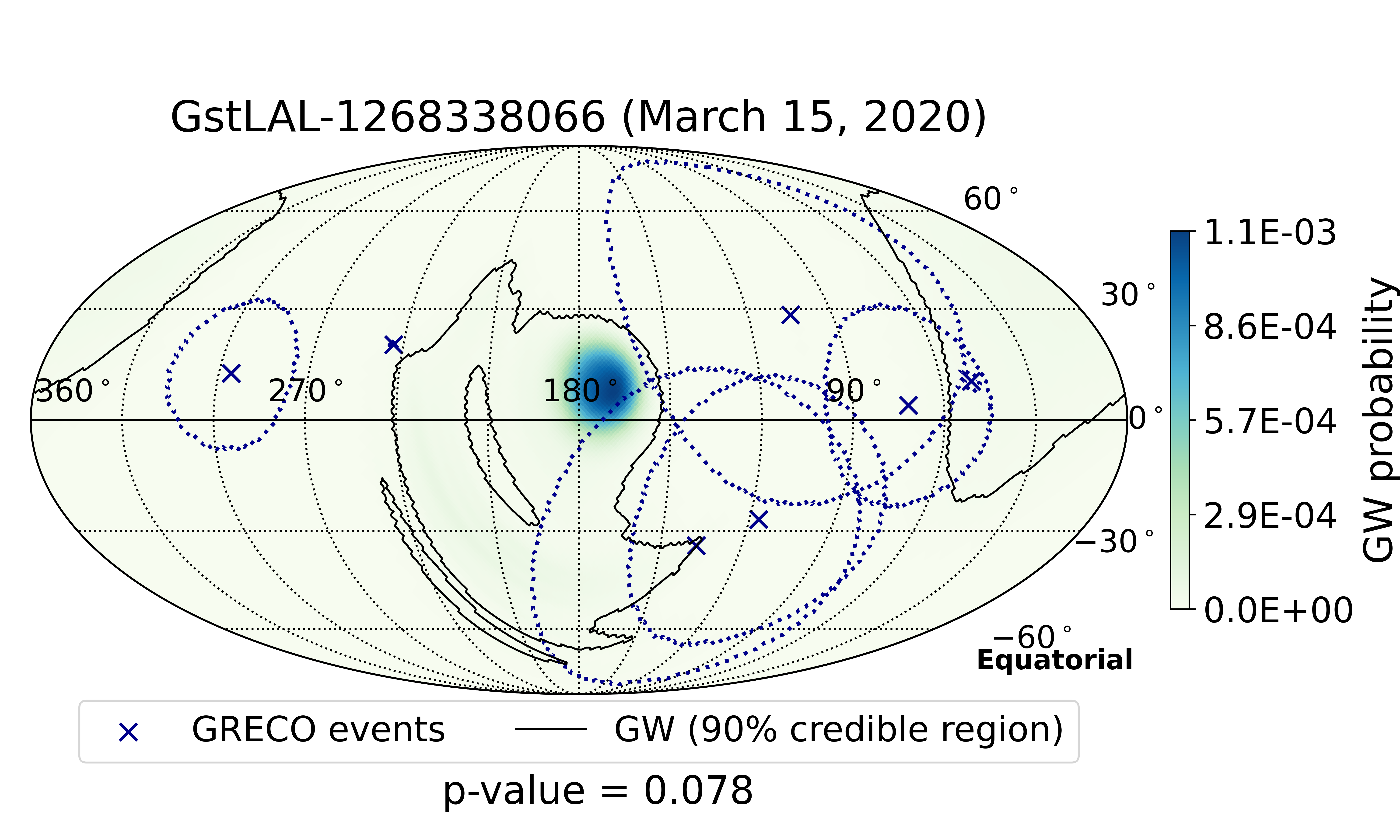}
    \includegraphics[width=0.49\linewidth]{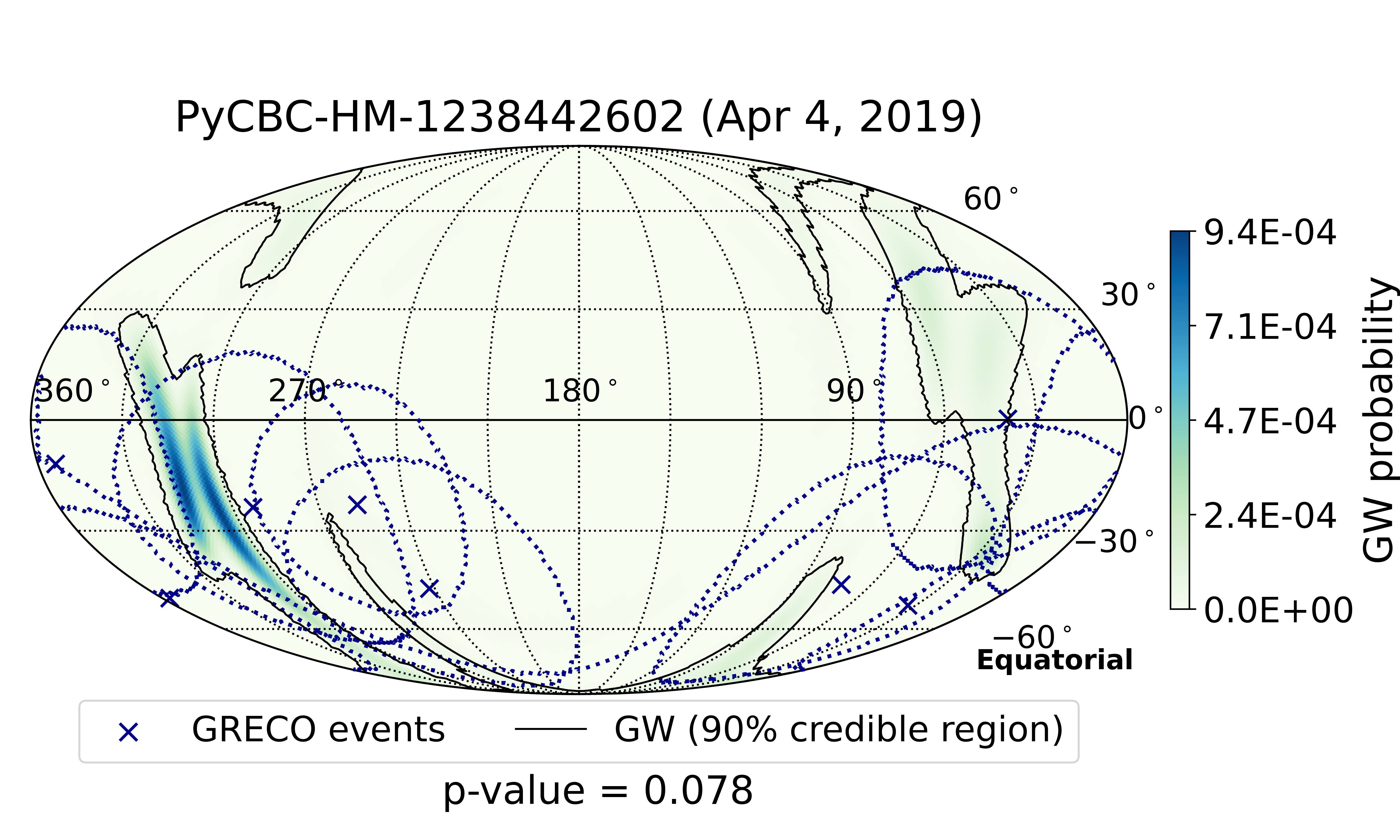}
    \caption{Joint GW and neutrino skymaps (continued).}
\end{figure}

\begin{figure}[htb!]\ContinuedFloat
    \centering
    \includegraphics[width=0.49\linewidth]{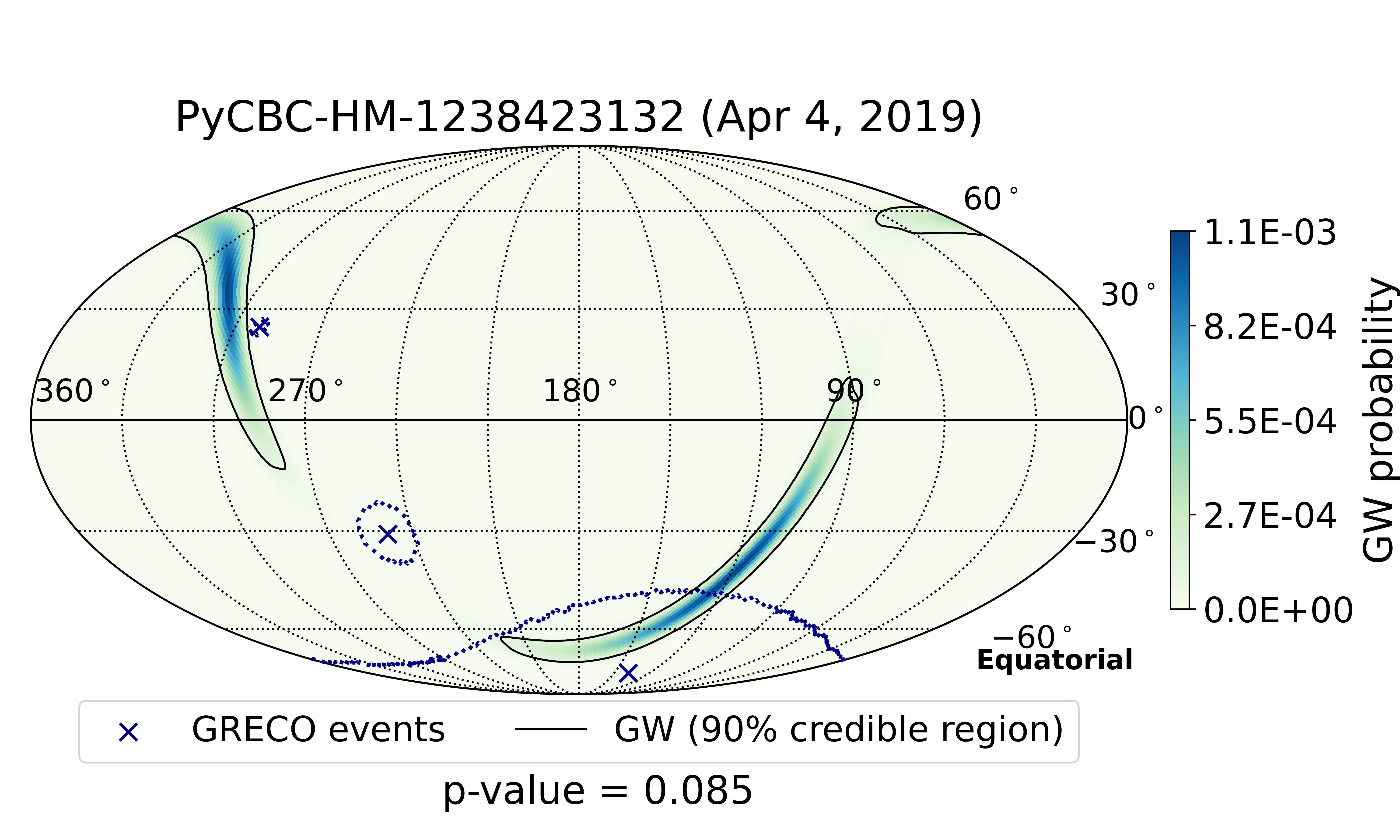}
    \includegraphics[width=0.49\linewidth]{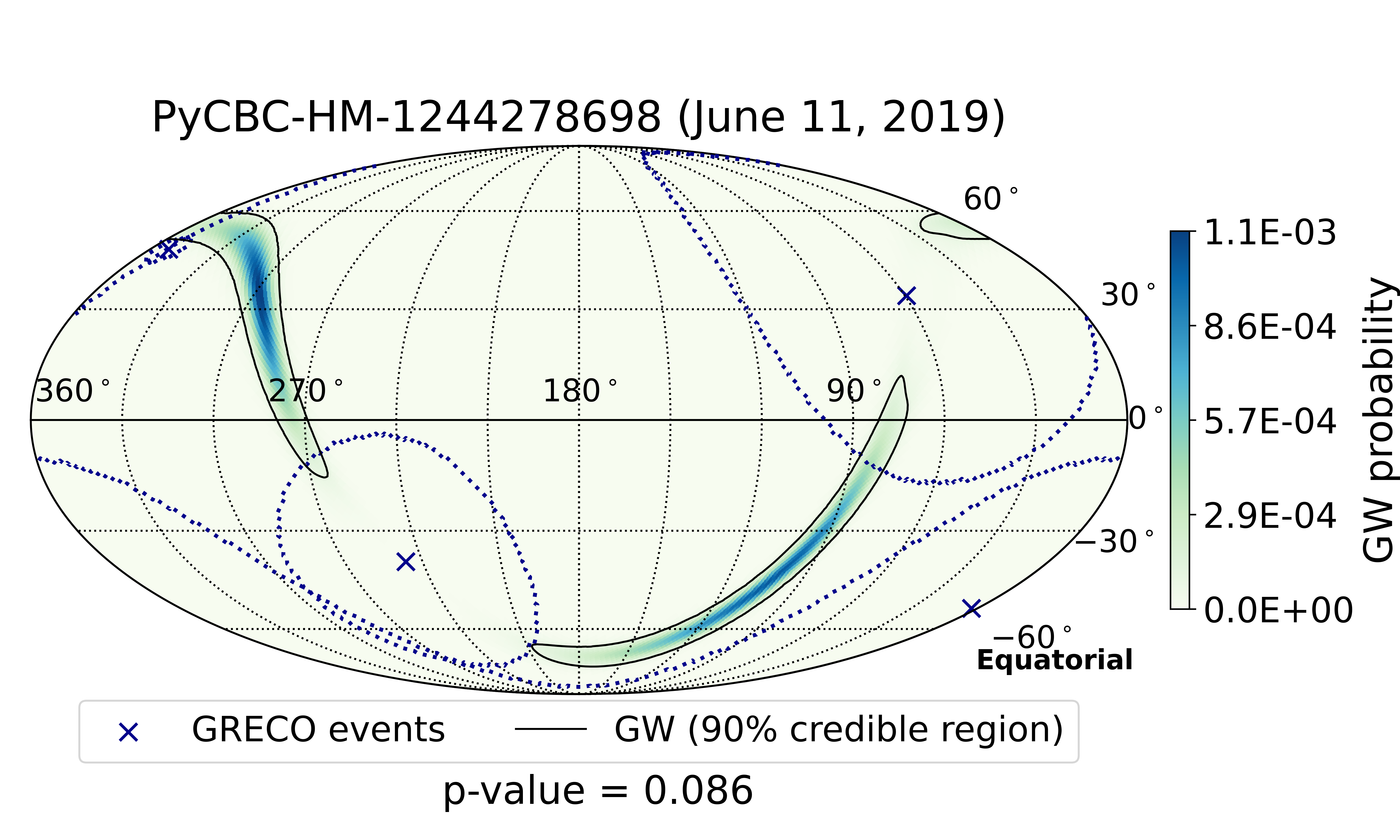}
    \includegraphics[width=0.49\linewidth]{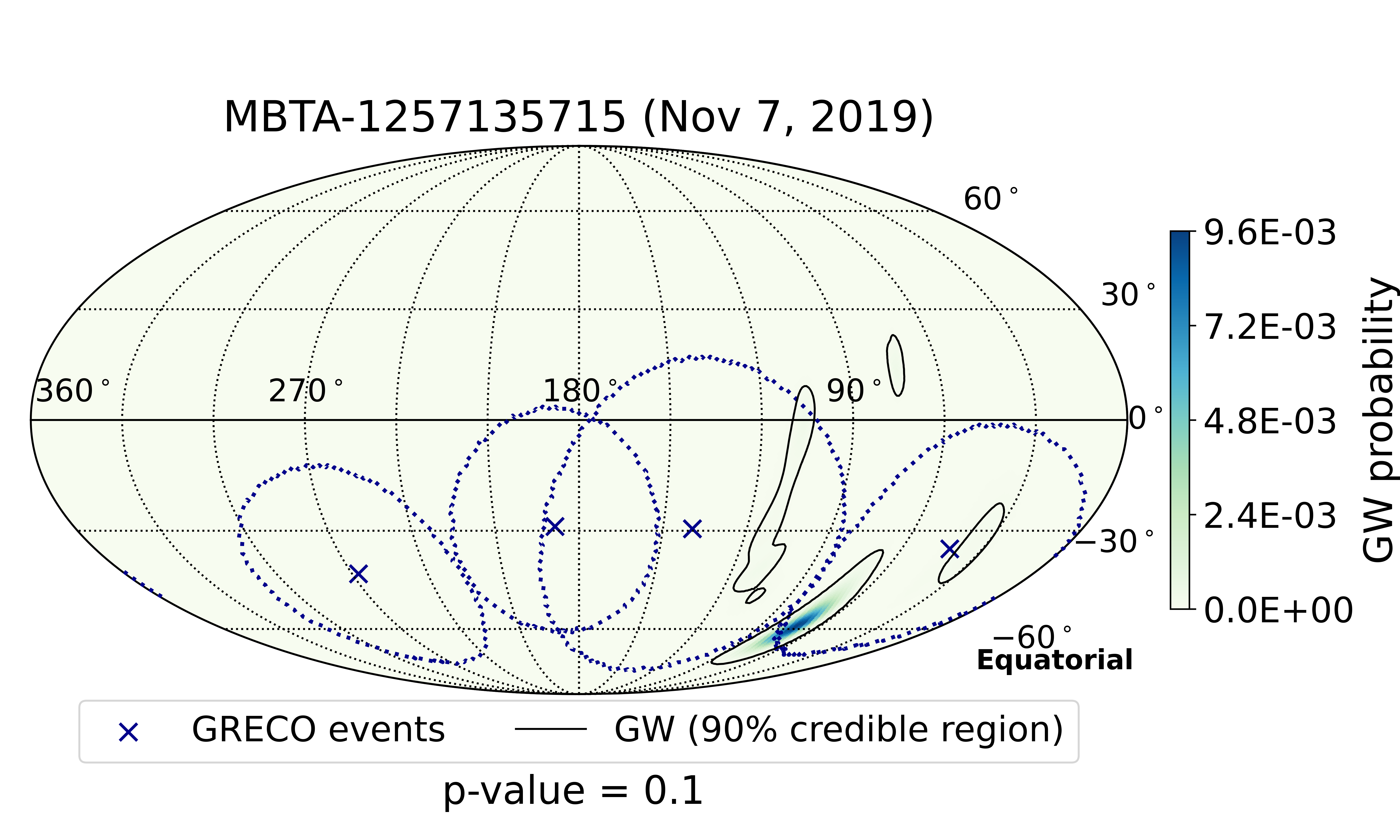}
    \includegraphics[width=0.49\linewidth]{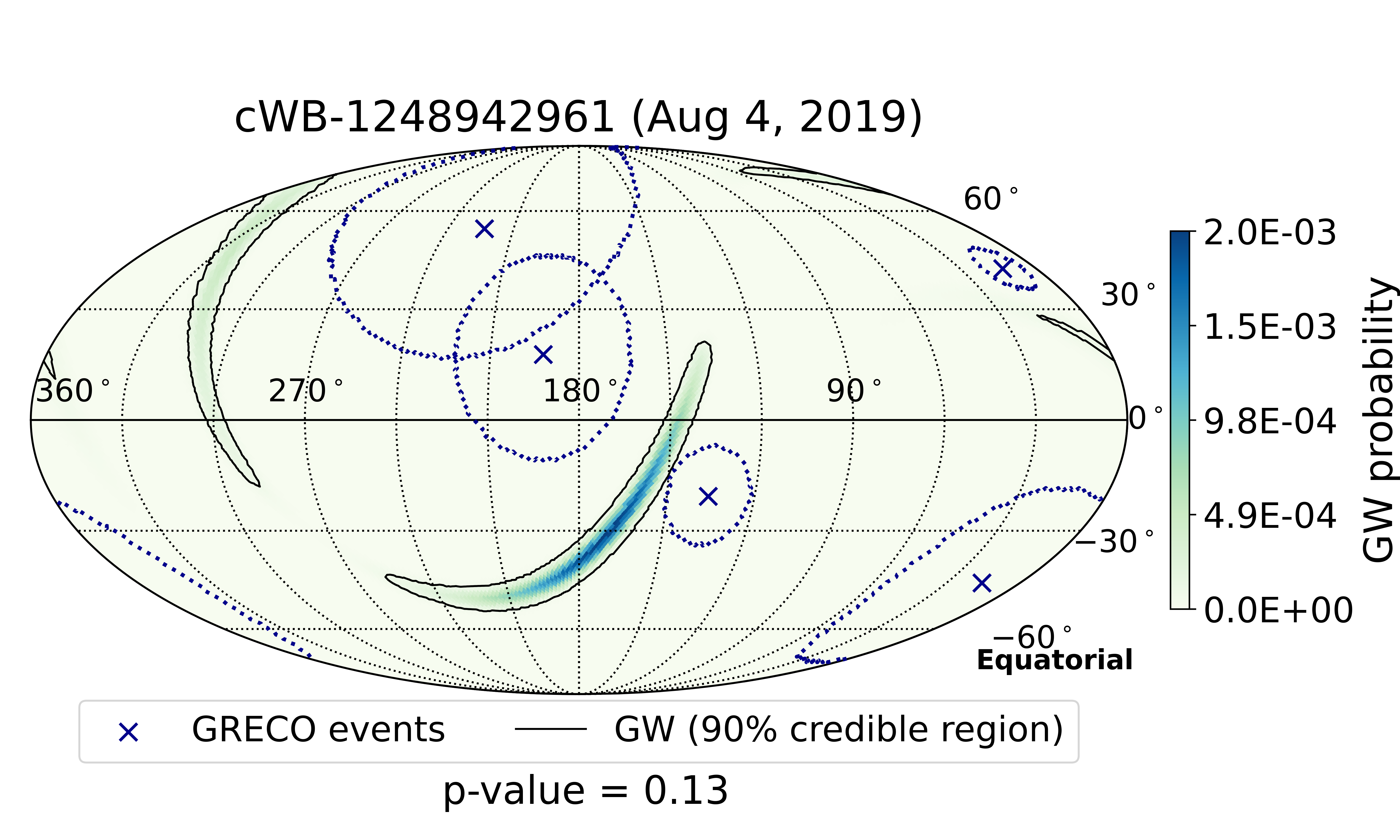}
    \includegraphics[width=0.49\linewidth]{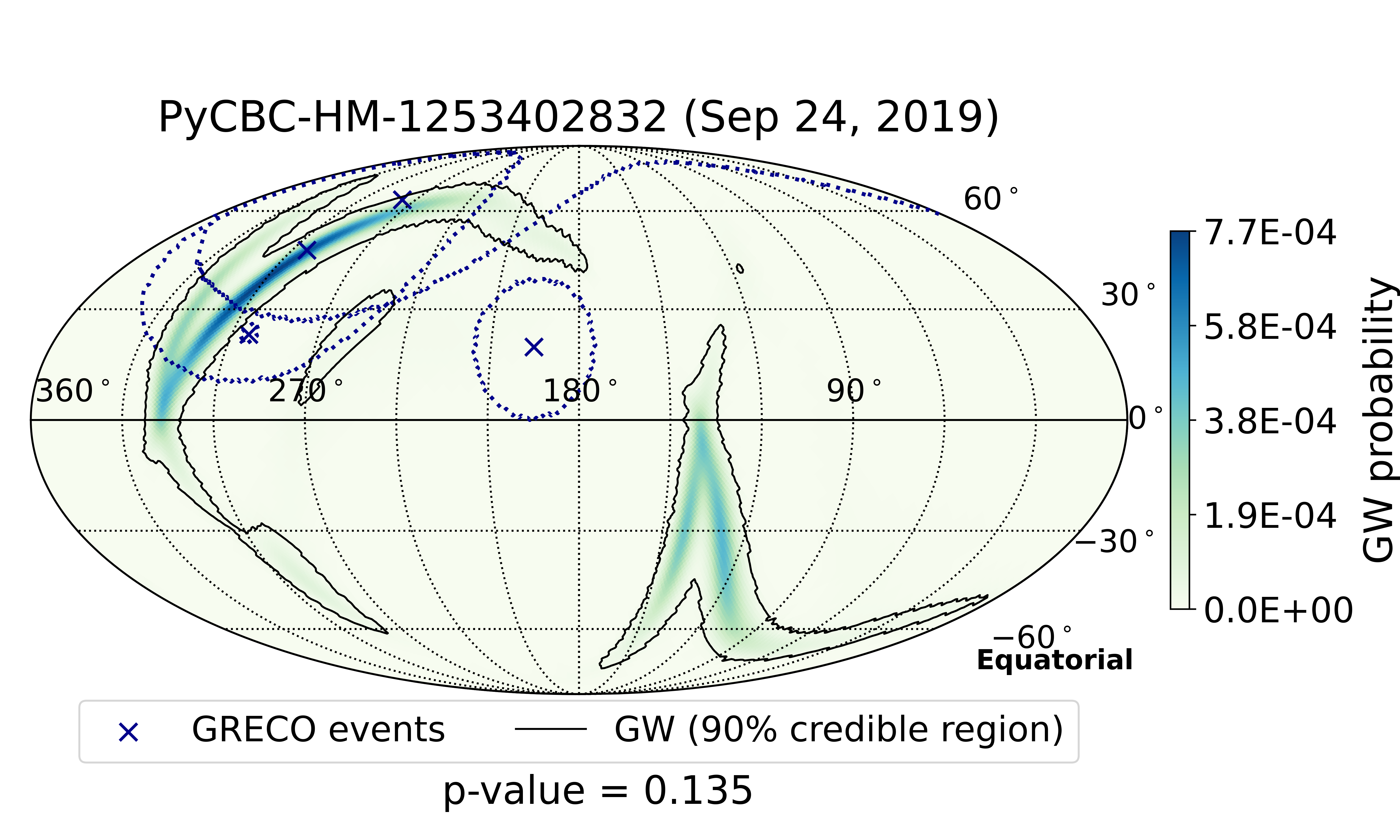}
    \caption{Joint GW + neutrino skymaps for the sub-threshold GW events from GWTC-2.1 and GWTC-3, for which the pre-trial p-value $\leq$ 0.14 ($\geq 1 \sigma$). The neutrino events marked with blue crosses overlaid on each GW skymap were observed within the 1000~s time window around the respective GW detection time. The skymaps are organised in descending order of significance for neutrino correlation, as shown in Tab. \ref{tab:unblinding_results}. The most significant candidate is not shown here, as it is separately discussed in the paper. The per-event pre-trial $p$-value has been annotated on the individual skymap.}
    \label{fig:unblinded_skymaps_most_sig}
\end{figure}

\begin{figure}[htb!]
    \centering
    \includegraphics[width=0.49\linewidth]{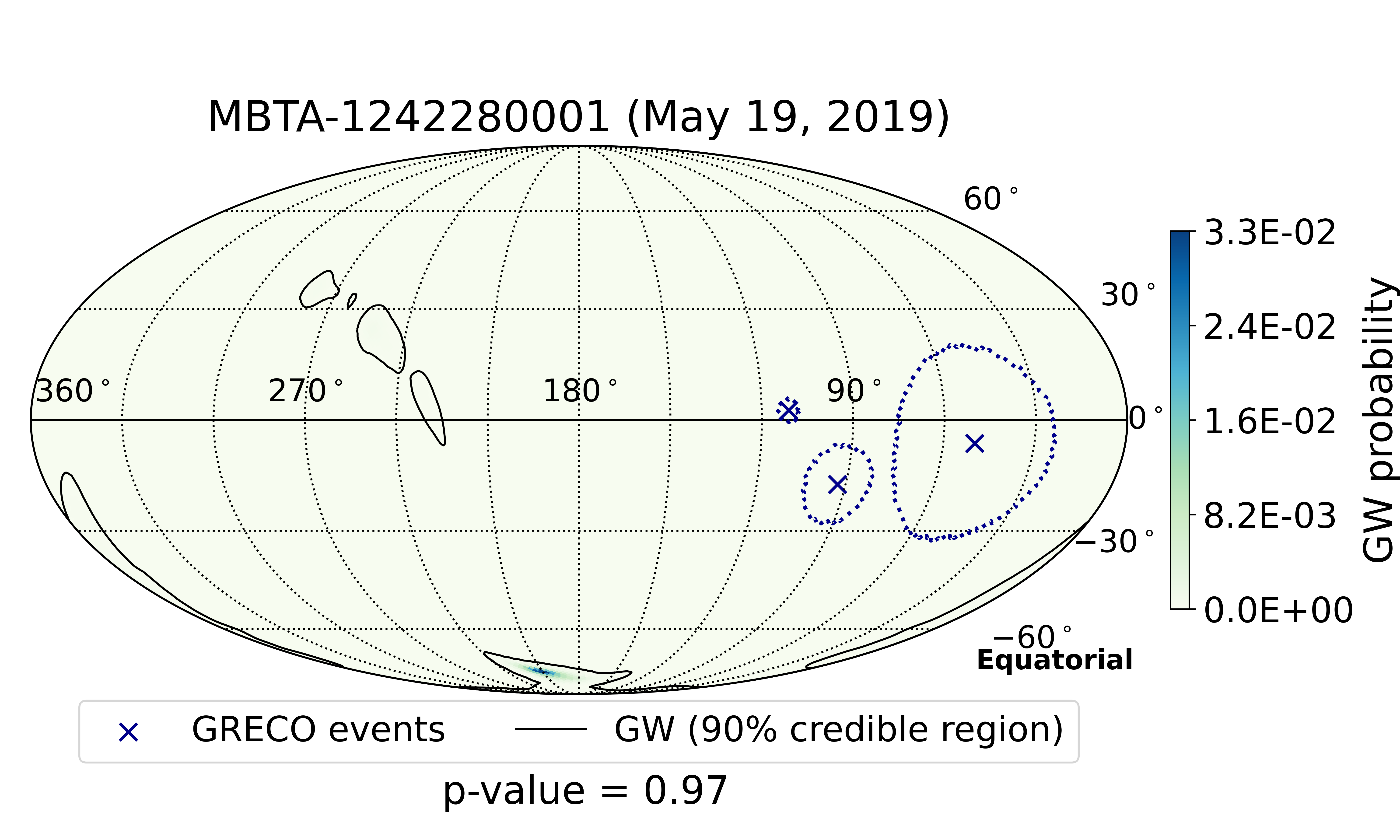}
    \includegraphics[width=0.49\linewidth]{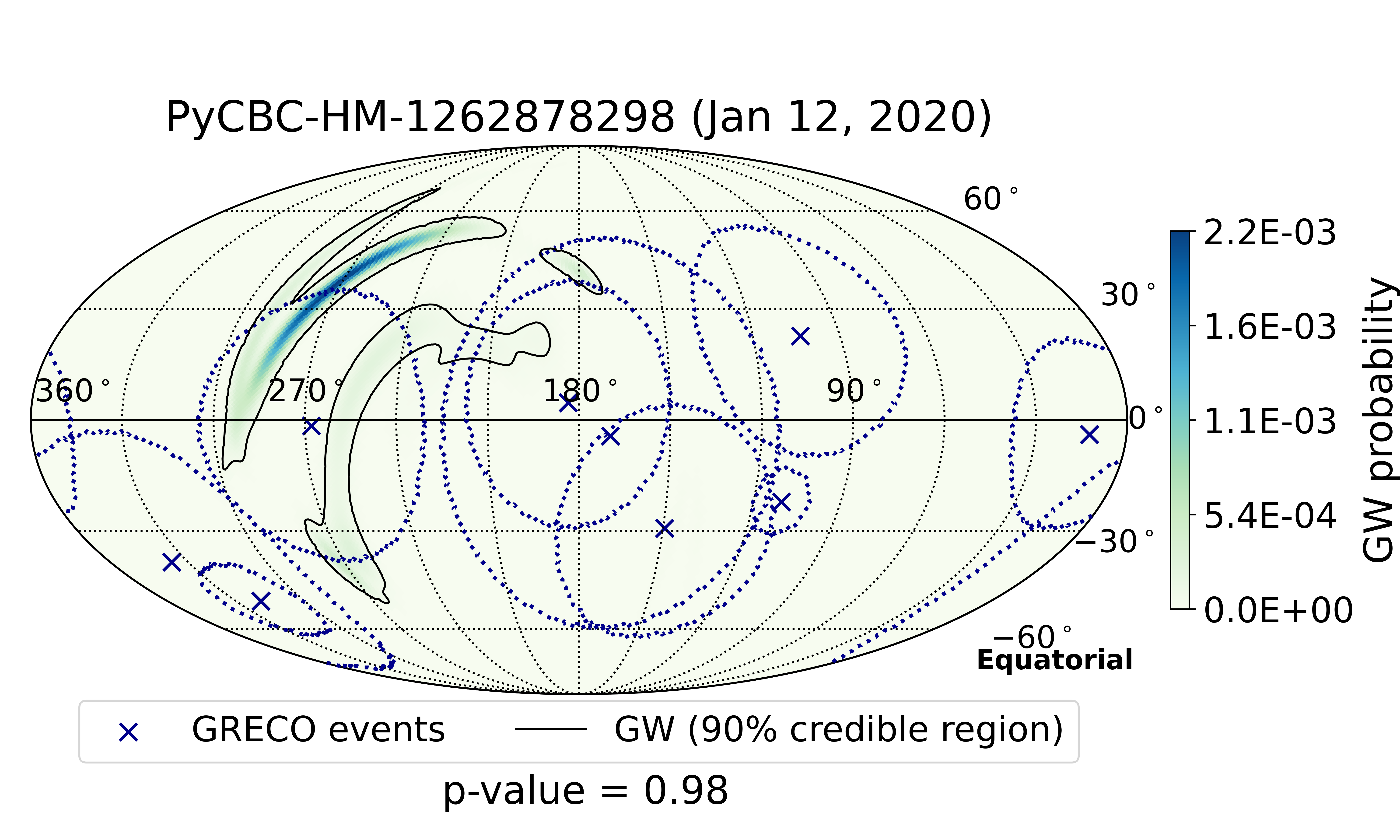}
    \includegraphics[width=0.49\linewidth]{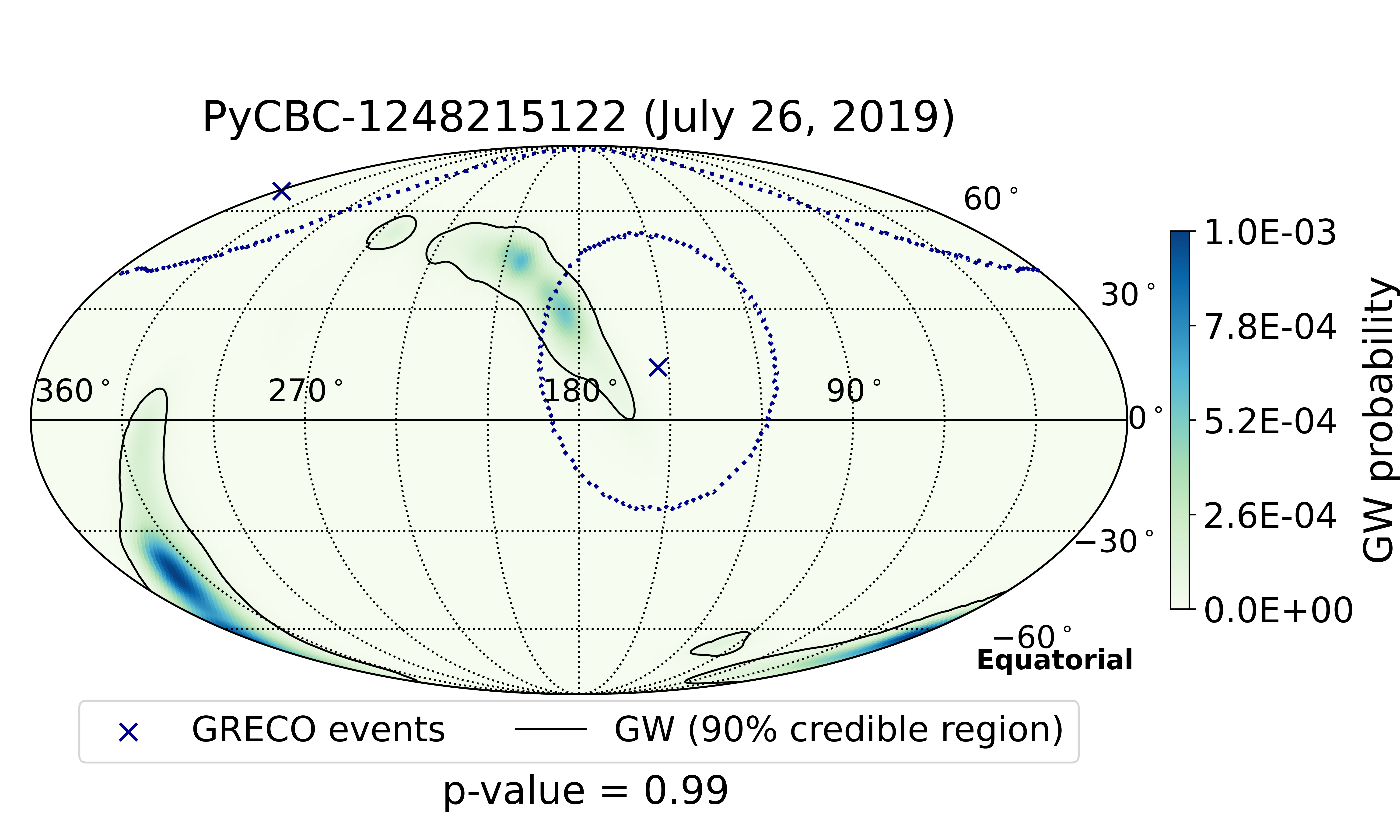}
    \caption{Joint GW + neutrino skymaps for three subthreshold GW events detected during O3, with highest pre-trial $p$-value ($>$ 0.96). The neutrino events marked with blue crosses overlaid on each GW skymap were observed within the 1000~s time window around the respective GW detection time. But none of the neutrino events claims a significant coincidence. The skymaps are organised in the descending order of significance for neutrino correlation, as shown in Tab. \ref{tab:unblinding_results}. The per-event pre-trial $p$-value has been annotated on the individual skymap.
}
    \label{fig:unblinded_skymaps_least_sig}
\end{figure}

Additionally, the skymaps for 3 GW events, which have the least correlation with the neutrino events detected during the 1000~s search time window, are shown in Fig. \ref{fig:unblinded_skymaps_least_sig}. They correspond to the highest pre-trial $p$-values obtained from the analysis. These are GW events with visibly no spatial coincidence with the detected neutrinos within the 1000~s time window, thus leading towards the highest probability of coincidence due to background fluctuations.
%This confirms that the temporally coincident neutrinos with no significant spatial association with the GW localisation regions lead to results consistent with background expectation. This shows the power of the analysis. 

%\section{Author publication charges} \label{sec:pubcharge}

%% For this sample we use BibTeX plus aasjournalv7.bst to generate the

\bibliography{bibliography}{}

@article{gwtc-1,
    author = "Abbott, B. P. and others",
    collaboration = "LIGO Scientific, Virgo",
    title = "{GWTC-1: A Gravitational-Wave Transient Catalog of Compact Binary Mergers Observed by LIGO and Virgo during the First and Second Observing Runs}",
    eprint = "1811.12907",
    archivePrefix = "arXiv",
    primaryClass = "astro-ph.HE",
    reportNumber = "LIGO-P1800307",
    doi = "10.1103/PhysRevX.9.031040",
    journal = "Phys. Rev. X",
    volume = "9",
    number = "3",
    pages = "031040",
    year = "2019"
}

@article{IC_Upgrade_sensitivity,
    author = "Abbasi, R. and others",
    title = "{Astrophysical Sensitivity Projections for the IceCube Upgrade}",
    eprint = "2608.28395",
    archivePrefix = "arXiv",
    primaryClass = "astro-ph.HE",
    month = "8",
    year = "2026"
}

@article{gwtc-2,
    author = "Abbott, R. and others",
    collaboration = "LIGO Scientific, Virgo",
    title = "{GWTC-2: Compact Binary Coalescences Observed by LIGO and Virgo During the First Half of the Third Observing Run}",
    eprint = "2010.14527",
    archivePrefix = "arXiv",
    primaryClass = "gr-qc",
    reportNumber = "P2000061",
    doi = "10.1103/PhysRevX.11.021053",
    journal = "Phys. Rev. X",
    volume = "11",
    pages = "021053",
    year = "2021"
}

@article{gwtc-3,
    author = "Abbott, R. and others",
    collaboration = "LIGO Scientific, the Virgo, and the KAGRA",
    title = "{GWTC-3: Compact Binary Coalescences Observed by LIGO and Virgo during the Second Part of the Third Observing Run}",
    eprint = "2111.03606",
    archivePrefix = "arXiv",
    primaryClass = "gr-qc",
    reportNumber = "LIGO-P2000318",
    doi = "10.1103/PhysRevX.13.041039",
    journal = "Phys. Rev. X",
    volume = "13",
    number = "4",
    pages = "041039",
    year = "2023"
}

@article{GW170817,
    author = "Abbott, B. P. and others",
    collaboration = "LIGO Scientific, Virgo, Fermi GBM, INTEGRAL, IceCube, AstroSat Cadmium Zinc Telluride Imager Team, IPN, Insight-Hxmt, ANTARES, Swift, AGILE Team, 1M2H Team, Dark Energy Camera GW-EM, DES, DLT40, GRAWITA, Fermi-LAT, ATCA, ASKAP, Las Cumbres Observatory Group, OzGrav, DWF (Deeper Wider Faster Program), AST3, CAASTRO, VINROUGE, MASTER, J-GEM, GROWTH, JAGWAR, CaltechNRAO, TTU-NRAO, NuSTAR, Pan-STARRS, MAXI Team, TZAC Consortium, KU, Nordic Optical Telescope, ePESSTO, GROND, Texas Tech University, SALT Group, TOROS, BOOTES, MWA, CALET, IKI-GW Follow-up, H.E.S.S., LOFAR, LWA, HAWC, Pierre Auger, ALMA, Euro VLBI Team, Pi of Sky, Chandra Team at McGill University, DFN, ATLAS Telescopes, High Time Resolution Universe Survey, RIMAS, RATIR, SKA South Africa/MeerKAT",
    title = "{Multi-messenger Observations of a Binary Neutron Star Merger}",
    eprint = "1710.05833",
    archivePrefix = "arXiv",
    primaryClass = "astro-ph.HE",
    reportNumber = "LIGO-P1700294, VIR-0802A-17, FERMILAB-PUB-17-478-A-AE-CD",
    doi = "10.3847/2041-8213/aa91c9",
    journal = "Astrophys. J. Lett.",
    volume = "848",
    number = "2",
    pages = "L12",
    year = "2017"
}

@article{TXS_IC,
  collaboration="{IceCube}",
  author = {Aartsen, Mark and others},
  title="{Neutrino emission from the direction of the blazar TXS 0506+ 056 prior to the IceCube-170922A alert}",
  journal={Science},
  volume={361},
  number={6398},
  pages={147--151},
  year={2018},
  publisher={American Association for the Advancement of Science},
  doi = "10.1126/science.aat2890"
}

@article{GW150914,
    author = "Abbott, B. P. and others",
    collaboration = "LIGO Scientific and the Virgo",
    title = "{Observation of Gravitational Waves from a Binary Black Hole Merger}",
    eprint = "1602.03837",
    archivePrefix = "arXiv",
    primaryClass = "gr-qc",
    reportNumber = "LIGO-P150914",
    doi = "10.1103/PhysRevLett.116.061102",
    journal = "Phys. Rev. Lett.",
    volume = "116",
    number = "6",
    pages = "061102",
    year = "2016"
}

@article{novae,
    author = {Abbasi, R. and others},
    collaboration = "IceCube",
    title = "{Search for sub-TeV Neutrino Emission from Novae with IceCube-DeepCore}",
    eprint = "2212.06810",
    archivePrefix = "arXiv",
    primaryClass = "astro-ph.HE",
    doi = "10.3847/1538-4357/acdc1b",
    journal = "Astrophys. J.",
    volume = "953",
    number = "2",
    pages = "160",
    year = "2023",
    note = "[Erratum: Astrophys.J. 971, (2024)]"
}

@article{cwb_allsky,
    author={Klimenko, Sergey and others},
title="{Method for Detection and Reconstruction of Gravitational Wave Transients with Networks of Advanced Detectors}",
    journal={Physical Review D},
    volume={93},
  number={4},
  pages={042004},
  year={2016},
  publisher={APS},
  doi = "10.1103/PhysRevD.93.042004"
}

@article{greco-gw,
    author = "Abbasi, R. and others",
    collaboration = "IceCube",
    title = "{A Search for IceCube Sub-TeV Neutrinos Correlated with Gravitational-wave Events Detected By LIGO/Virgo}",
    eprint = "2303.15970",
    archivePrefix = "arXiv",
    primaryClass = "astro-ph.HE",
    doi = "10.3847/1538-4357/aceefc",
    journal = "Astrophys. J.",
    volume = "959",
    number = "2",
    pages = "96",
    year = "2023",
    note = "[Erratum: Astrophys.J. 971, 192 (2024)]"
}

@article{greco_Erratum,
  author = "Abbasi, R. and others",
  collaboration = "IceCube",
  title   = "{Erratum: "A Search for IceCube Sub-TeV Neutrinos Correlated with Gravitational-wave Events Detected By LIGO/Virgo" (2023, ApJ, 959, 96)}",
  doi     = "10.3847/1538-4357/ad683f",
  journal = "Astrophys. J.",
  volume  = "971",
  number  = "2",
  pages   = "192",
  year    = "2024"
}

@article{gstlal,
  title= "{GstLAL: A software framework for gravitational wave discovery}",
  author={Cannon, Kipp and others},
  journal={SoftwareX},
  volume={14},
  pages={100680},
  year={2021},
  doi = {10.1016/j.softx.2021.100680},
  publisher={Elsevier}
}

@article{mbta,
    author = "All{\'e}n{\'e}, Christopher and others",
    title = "{The MBTA pipeline for detecting compact binary coalescences in the fourth LIGO-Virgo-KAGRA observing run}",
    eprint = "2501.04598",
    archivePrefix = "arXiv",
    primaryClass = "gr-qc",
    doi = "10.1088/1361-6382/add234",
    journal = "Class. Quant. Grav.",
    volume = "42",
    number = "10",
    pages = "105009",
    year = "2025"
}

@article{pycbc,
    author = "Usman, Samantha A. and others",
    title = "{The PyCBC search for gravitational waves from compact binary coalescence}",
    eprint = "1508.02357",
    archivePrefix = "arXiv",
    primaryClass = "gr-qc",
    reportNumber = "LIGO-P1500086",
    doi = "10.1088/0264-9381/33/21/215004",
    journal = "Class. Quant. Grav.",
    volume = "33",
    number = "21",
    pages = "215004",
    year = "2016"
}

@article{pycbc-HM,
    author = "Nitz, Alexander Harvey",
    title = "{Distinguishing short duration noise transients in LIGO data to improve the PyCBC search for gravitational waves from high mass binary black hole mergers}",
    eprint = "1709.08974",
    archivePrefix = "arXiv",
    primaryClass = "gr-qc",
    reportNumber = "LIGO-P1700275",
    doi = "10.1088/1361-6382/aaa13d",
    journal = "Class. Quant. Grav.",
    volume = "35",
    number = "3",
    pages = "035016",
    year = "2018"
}

@article{deepcore,
    author = "Abbasi, R. and others",
    collaboration = "IceCube",
    title = "{The Design and Performance of IceCube DeepCore}",
    eprint = "1109.6096",
    archivePrefix = "arXiv",
    primaryClass = "astro-ph.IM",
    doi = "10.1016/j.astropartphys.2012.01.004",
    journal = "Astropart. Phys.",
    volume = "35",
    pages = "615--624",
    year = "2012"
}

@article{GWfollowupO3,
    author = "Abbasi, R. and others",
    collaboration = "IceCube",
    title = "{IceCube Search for Neutrinos Coincident with Gravitational Wave Events from LIGO/Virgo Run O3}",
    eprint = "2208.09532",
    archivePrefix = "arXiv",
    primaryClass = "astro-ph.HE",
    doi = "10.3847/1538-4357/aca5fc",
    journal = "Astrophys. J.",
    volume = "944",
    number = "1",
    pages = "80",
    year = "2023"
}

@article{ICRC23_proceeding,
    author = "Mukherjee, Tista and others",
    collaboration = "IceCube",
    title = "{Searching for IceCube sub-TeV neutrino counterparts to sub-threshold Gravitational Wave events}",
    eprint = "2308.06102",
    archivePrefix = "arXiv",
    primaryClass = "astro-ph.HE",
    reportNumber = "PoS-ICRC2023-1504",
    doi = "10.22323/1.444.1504",
    journal = "PoS",
    volume = "ICRC2023",
    pages = "1504",
    year = "2023"
}

@article{Biehl_offAxisGRB,
    author = "Biehl, Daniel and Heinze, Jonas and Winter, Walter",
    title = "{Expected neutrino fluence from short Gamma-Ray Burst 170817A and off-axis angle constraints}",
    eprint = "1712.00449",
    archivePrefix = "arXiv",
    primaryClass = "astro-ph.HE",
    reportNumber = "DESY-17-230",
    doi = "10.1093/mnras/sty285",
    journal = "Mon. Not. Roy. Astron. Soc.",
    volume = "476",
    number = "1",
    pages = "1191--1197",
    year = "2018"
}

@article{elevated_pastro,
    author = "Veske, Do{\u{g}}a and Zhang, Albert and M{\'a}rka, Zsuzsa and M{\'a}rka, Szabolcs",
    title = "{Enhancing the significance of astrophysical events with multimessenger coincidences}",
    eprint = "2407.08732",
    archivePrefix = "arXiv",
    primaryClass = "astro-ph.HE",
    doi = "10.1103/PhysRevD.111.103013",
    journal = "Phys. Rev. D",
    volume = "111",
    number = "10",
    pages = "103013",
    year = "2025"
}

@article{IC_AstroNu_2013,
    author = "Aartsen, M. G. and others",
    collaboration = "IceCube",
    title = "{Evidence for High-Energy Extraterrestrial Neutrinos at the IceCube Detector}",
    eprint = "1311.5238",
    archivePrefix = "arXiv",
    primaryClass = "astro-ph.HE",
    doi = "10.1126/science.1242856",
    journal = "Science",
    volume = "342",
    pages = "1242856",
    year = "2013"
}

@article{IC_JINST,
    author = "Aartsen, M. G. and others",
    collaboration = "IceCube",
    title = "{The IceCube Neutrino Observatory: Instrumentation and Online Systems}",
    eprint = "1612.05093",
    archivePrefix = "arXiv",
    primaryClass = "astro-ph.IM",
    doi = "10.1088/1748-0221/12/03/P03012",
    journal = "JINST",
    volume = "12",
    number = "03",
    pages = "P03012",
    year = "2017",
    note = "[Erratum: JINST 19, E05001 (2024)]"
}

@article{ECRS24_TM,
    collaboration = "{IceCube}",    
    title = "{Searching for sub-TeV IceCube neutrinos correlated to sub-threshold GW events}",
    author = {Mukherjee, Tista and others},
    journal={arXiv preprint arXiv:2506.09694},
    year = {2025},
    doi="10.48550/arXiv.2506.09694"
}

@article{GWHENsearch_GW170817,
    author = "Albert, A. and others",
    collaboration = "ANTARES, IceCube, Pierre Auger, LIGO Scientific, Virgo",
    title = "{Search for High-energy Neutrinos from Binary Neutron Star Merger GW170817 with ANTARES, IceCube, and the Pierre Auger Observatory}",
    eprint = "1710.05839",
    archivePrefix = "arXiv",
    primaryClass = "astro-ph.HE",
    reportNumber = "LIGO-P1700344, FERMILAB-PUB-17-471-ND-TD, LIGO-P1700344",
    doi = "10.3847/2041-8213/aa9aed",
    journal = "Astrophys. J. Lett.",
    volume = "850",
    number = "2",
    pages = "L35",
    year = "2017"
}

@article{NGC1068_IC,
    author = "Abbasi, R. and others",
    collaboration = "IceCube",
    title = "{Evidence for neutrino emission from the nearby active galaxy NGC 1068}",
    eprint = "2211.09972",
    archivePrefix = "arXiv",
    primaryClass = "astro-ph.HE",
    doi = "10.1126/science.abg3395",
    journal = "Science",
    volume = "378",
    number = "6619",
    pages = "538--543",
    year = "2022"
}

@article{GP_IC,
    author = "Abbasi, R. and others",
    collaboration = "IceCube",
    title = "{Observation of high-energy neutrinos from the Galactic plane}",
    eprint = "2307.04427",
    archivePrefix = "arXiv",
    primaryClass = "astro-ph.HE",
    doi = "10.1126/science.adc9818",
    journal = "Science",
    volume = "380",
    number = "6652",
    pages = "adc9818",
    year = "2023"
}

@article{GWTC1_ICfollowup,
    author = "Aartsen, M. G. and others",
    collaboration = "IceCube",
    title = "{IceCube Search for Neutrinos Coincident with Compact Binary Mergers from LIGO-Virgo{\textquoteright}s First Gravitational-wave Transient Catalog}",
    eprint = "2004.02910",
    archivePrefix = "arXiv",
    primaryClass = "astro-ph.HE",
    doi = "10.3847/2041-8213/ab9d24",
    journal = "Astrophys. J. Lett.",
    volume = "898",
    number = "1",
    pages = "L10",
    year = "2020"
}

@article{gw_nu,
    author = "Balagopal V., Aswathi and others",
    collaboration = "IceCube",
    title = "{Searches for IceCube Neutrinos Coincident with Gravitational Wave Events}",
    doi = "10.22323/1.444.1484",
    journal = "PoS",
    volume = "ICRC2023",
    pages = "1484",
    year = "2023"
}

@inproceedings{ICRC25_aswathi,
    author = "Balagopal V., Aswathi and Hori, Sam and Vandenbroucke, Justin",
    title = "{Multi-Energy and Multi-Sample Searches for Neutrinos from GW Events}",
    booktitle = "{39th International Cosmic Ray Conference}",
    eprint = "2508.09034",
    archivePrefix = "arXiv",
    primaryClass = "astro-ph.HE",
    reportNumber = "PoS-ICRC2025-915",
    year = "2025"
}

@inproceedings{GFU_O4follow_up,
    author = "Marka, Zsuzsa and Thwaites, Jessie and Vandenbroucke, Justin and Veske, Do{\u{g}}a",
    title = "{Results from IceCube Searches for High-energy Neutrinos Coincident with Gravitational-Wave Alerts in LVK O4}",
    booktitle = "{39th International Cosmic Ray Conference}",
    eprint = "2507.08131",
    archivePrefix = "arXiv",
    primaryClass = "astro-ph.HE",
    reportNumber = "PoS-ICRC2025-1113",
    year = "2025"
}

@article{LVKO4a_realTime_GFU,
    author = "Abbasi, R. and others",
    collaboration = "IceCube",
    title = "{IceCube Real-time Searches for High-energy Neutrinos Coincident with LIGO/Virgo/KAGRA Gravitational-Wave Alerts in O4a}",
    eprint = "2606.13762",
    archivePrefix = "arXiv",
    primaryClass = "astro-ph.HE",
    month = "6",
    year = "2026"
}

@article{GWfollowUp_MeVnu,
    author = "Valtonen-Mattila, Nora",
    title = "{IceCube Search for MeV Neutrinos from Mergers using Gravitational Wave Catalogs}",
    eprint = "2603.13076",
    archivePrefix = "arXiv",
    primaryClass = "astro-ph.HE",
    month = "3",
    year = "2026"
}

@article{IC_GRB_2022,
    author = "Abbasi, R. and others",
    title = "{Searches for Neutrinos from Gamma-Ray Bursts Using the IceCube Neutrino Observatory}",
    eprint = "2205.11410",
    archivePrefix = "arXiv",
    primaryClass = "astro-ph.HE",
    doi = "10.3847/1538-4357/ac9785",
    journal = "Astrophys. J.",
    volume = "939",
    number = "2",
    pages = "116",
    year = "2022"
}

@article{GRECO_3yr_transient,
    author = "Abbasi, R. and others",
    collaboration = "IceCube",
    title = "{First all-flavor search for transient neutrino emission using 3-years of IceCube DeepCore data}",
    eprint = "2011.05096",
    archivePrefix = "arXiv",
    primaryClass = "astro-ph.HE",
    doi = "10.1088/1475-7516/2022/01/027",
    journal = "JCAP",
    volume = "01",
    pages = "027",
    year = "2022"
}

@article{greco_grb,
    author = "Abbasi, R. and others",
    collaboration = "IceCube",
    title = "{Search for 10{\textendash}1000 GeV Neutrinos from Gamma-Ray Bursts with IceCube}",
    eprint = "2312.11515",
    archivePrefix = "arXiv",
    primaryClass = "astro-ph.HE",
    doi = "10.3847/1538-4357/ad220b",
    journal = "Astrophys. J.",
    volume = "964",
    number = "2",
    pages = "126",
    year = "2024"
}

@article{Biswa_AGNdisk_BBH_2025,
    author = "Joshi, Raj Kishor and Bhake, Aryan and Banerjee, Biswajit and Vaidya, Bhargav and Ruiz, Milton and Tsokaros, Antonios and Mignone, Andrea and Branchesi, Marica and Shukla, Amit and {\v{C}}emelji{\'c}, Miljenko",
    title = "{Binary black holes in magnetized disks of active galactic nuclei}",
    eprint = "2509.16796",
    archivePrefix = "arXiv",
    primaryClass = "astro-ph.HE",
    doi = "10.1051/0004-6361/202555874",
    journal = "Astron. Astrophys.",
    volume = "703",
    pages = "A304",
    year = "2025"
}

@article{Bartos_BBH_AGN,
    author = "Bartos, Imre and Kocsis, Bence and Haiman, Zolt and M{\'a}rka, Szabolcs",
    title = "{Rapid and Bright Stellar-mass Binary Black Hole Mergers in Active Galactic Nuclei}",
    eprint = "1602.03831",
    archivePrefix = "arXiv",
    primaryClass = "astro-ph.HE",
    doi = "10.3847/1538-4357/835/2/165",
    journal = "Astrophys. J.",
    volume = "835",
    number = "2",
    pages = "165",
    year = "2017"
}

@article{BBH_AGN_MNRAS,
    author = "Zhu, Jin-Ping",
    title = "{High-energy neutrinos from merging stellar-mass black holes in active galactic nuclei accretion disc}",
    eprint = "2310.14255",
    archivePrefix = "arXiv",
    primaryClass = "astro-ph.HE",
    doi = "10.1093/mnrasl/slad176",
    journal = "Mon. Not. Roy. Astron. Soc.",
    volume = "528",
    number = "1",
    pages = "L88--L95",
    year = "2023"
}

@article{BBH_AGN_APJL,
    author = "Zhou, Zi-Hang and Wang, Kai",
    title = "{High-energy Neutrino Emission Associated with GWs from Binary Black Hole Mergers in AGN Accretion Disks}",
    eprint = "2310.15832",
    archivePrefix = "arXiv",
    primaryClass = "astro-ph.HE",
    doi = "10.3847/2041-8213/ad096f",
    journal = "Astrophys. J. Lett.",
    volume = "958",
    number = "1",
    pages = "L12",
    year = "2023"
}

@article{ZTF_BBH_kilonova,
    author = "Graham, M. J. and others",
    title = "{Candidate Electromagnetic Counterpart to the Binary Black Hole Merger Gravitational Wave Event S190521g}",
    eprint = "2006.14122",
    archivePrefix = "arXiv",
    primaryClass = "astro-ph.HE",
    doi = "10.1103/PhysRevLett.124.251102",
    journal = "Phys. Rev. Lett.",
    volume = "124",
    number = "25",
    pages = "251102",
    year = "2020"
}

@article{GW190521,
    author = "Abbott, R. and others",
    collaboration = "LIGO Scientific, Virgo",
    title = "{GW190521: A Binary Black Hole Merger with a Total Mass of $150  M_{\odot}$}",
    eprint = "2009.01075",
    archivePrefix = "arXiv",
    primaryClass = "gr-qc",
    doi = "10.1103/PhysRevLett.125.101102",
    journal = "Phys. Rev. Lett.",
    volume = "125",
    number = "10",
    pages = "101102",
    year = "2020"
}

@article{jointGWHENsearch_O1,
    author = "Albert, A. and others",
    collaboration = "ANTARES, IceCube, LIGO, Virgo",
    title = "{Search for Multimessenger Sources of Gravitational Waves and High-energy Neutrinos with Advanced LIGO during Its First Observing Run, ANTARES, and IceCube}",
    eprint = "1810.10693",
    archivePrefix = "arXiv",
    primaryClass = "astro-ph.HE",
    doi = "10.3847/1538-4357/aaf21d",
    journal = "Astrophys. J.",
    volume = "870",
    number = "2",
    pages = "134",
    year = "2019"
}

@article{Veske_subthGW_GFU,
    author = "Abbasi, R. and others",
    collaboration = "IceCube, LIGO Scientific, VIRGO, KAGRA",
    title = "{Deep Search for Joint Sources of Gravitational Waves and High-energy Neutrinos with IceCube during the Third Observing Run of LIGO and Virgo}",
    eprint = "2601.07595",
    archivePrefix = "arXiv",
    primaryClass = "astro-ph.HE",
    doi = "10.3847/1538-4357/ae4e1b",
    journal = "Astrophys. J.",
    volume = "1003",
    number = "1",
    pages = "41",
    year = "2026"
}

@article{sidak_correction,
    title={Rectangular confidence regions for the means of multivariate normal distributions},
    author={{\v{S}}id{\'a}k, Zbyn{\v{e}}k},
    doi = "10.2307/2283989",
    journal="{Journal of the American Statistical Association}",
    volume={62},
    number={318},
    pages={626--633},
    year={1967},
    publisher={Taylor \& Francis}
}

@article{IC_upgrade,
  title="{The IceCube Upgrade--Design and Science Goals}",
  author={Ishihara, Aya},
  journal={arXiv preprint arXiv:1908.09441},
  year={2019}
}

@article{Auger_UHEnu_BBHmerger,
    author = "Schimp, Michael and others",
    collaboration = "Pierre Auger",
    title = "{Combined Search for UHE Neutrinos from Binary Black Hole Mergers with the Pierre Auger Observatory}",
    doi = "10.22323/1.395.0968",
    journal = "PoS",
    volume = "ICRC2021",
    pages = "968",
    year = "2021"
}

@article{ANTARES_GWfollowUp_O3,
    author = "Albert, A. and others",
    collaboration = "ANTARES",
    title = "{Search for neutrino counterparts to the gravitational wave sources from LIGO/Virgo O3 run with the ANTARES detector}",
    eprint = "2302.07723",
    archivePrefix = "arXiv",
    primaryClass = "astro-ph.HE",
    doi = "10.1088/1475-7516/2023/04/004",
    journal = "JCAP",
    volume = "04",
    pages = "004",
    year = "2023"
}

@article{CBC_GRB_timeWindow,
    author = "Baret, Bruny and others",
    title = "{Bounding the Time Delay between High-energy Neutrinos and Gravitational-wave Transients from Gamma-ray Bursts}",
    eprint = "1101.4669",
    archivePrefix = "arXiv",
    primaryClass = "astro-ph.HE",
    doi = "10.1016/j.astropartphys.2011.04.001",
    journal = "Astropart. Phys.",
    volume = "35",
    pages = "1--7",
    year = "2011"
}

@article{EccentricBBH_2019,
    author = "Abbott, B. P. and others",
    collaboration = "LIGO Scientific, Virgo",
    title = "{Search for Eccentric Binary Black Hole Mergers with Advanced LIGO and Advanced Virgo during their First and Second Observing Runs}",
    eprint = "1907.09384",
    archivePrefix = "arXiv",
    primaryClass = "astro-ph.HE",
    reportNumber = "LIGO Document P1900110",
    doi = "10.3847/1538-4357/ab3c2d",
    journal = "Astrophys. J.",
    volume = "883",
    number = "2",
    pages = "149",
    year = "2019"
}

@article{EccentricBBH_O3,
    author = "Abac, A. G. and others",
    collaboration = "LIGO Scientific, KAGRA, VIRGO",
    title = "{Search for Eccentric Black Hole Coalescences during the Third Observing Run of LIGO and Virgo}",
    eprint = "2308.03822",
    archivePrefix = "arXiv",
    primaryClass = "astro-ph.HE",
    reportNumber = "LIGO-P2300080",
    doi = "10.3847/1538-4357/ad65ce",
    journal = "Astrophys. J.",
    volume = "973",
    number = "2",
    pages = "132",
    year = "2024"
}

@article{Healpix,
    author = "G{\'o}rski, K. M. and Hivon, E. and Banday, A. J. and Wandelt, B. D. and Hansen, F. K. and Reinecke, M. and Bartelmann, M.",
    title = "{HEALPix - A Framework for high resolution discretization, and fast analysis of data distributed on the sphere}",
    eprint = "astro-ph/0409513",
    archivePrefix = "arXiv",
    doi = "10.1086/427976",
    journal = "Astrophys. J.",
    volume = "622",
    pages = "759--771",
    year = "2005"
}

@article{ET_BlueBook,
    author = "Abac, Adrian and others",
    collaboration = "ET",
    title = "{The Science of the Einstein Telescope}",
    eprint = "2503.12263",
    archivePrefix = "arXiv",
    primaryClass = "gr-qc",
    reportNumber = "ET-0036C-25",
    doi = "10.1088/1475-7516/2026/03/081",
    journal = "JCAP",
    volume = "03",
    pages = "081",
    year = "2026"
}

@article{KM3NeT_O3followUp,
    author = "Aiello, S. and others",
    collaboration = "KM3NeT",
    title = "{Searches for neutrino counterparts of gravitational waves from the LIGO/Virgo third observing run with KM3NeT}",
    eprint = "2311.03804",
    archivePrefix = "arXiv",
    primaryClass = "astro-ph.HE",
    doi = "10.1088/1475-7516/2024/04/026",
    journal = "JCAP",
    volume = "04",
    pages = "026",
    year = "2024"
}

@article{SK_O4_GWfollowUp,
    author = "Machado, L. N.",
    collaboration = "Super-Kamiokande",
    title = "{Search for Neutrino Counterparts to Gravitational Wave Events in Super-Kamiokande during the LIGO/Virgo/KAGRA O4 Observing Run}",
    eprint = "2509.12002",
    archivePrefix = "arXiv",
    primaryClass = "hep-ex",
    doi = "10.1088/1742-6596/3177/1/012122",
    journal = "J. Phys. Conf. Ser.",
    volume = "3177",
    number = "1",
    pages = "012122",
    year = "2026"
}

@article{GW170817_BaikalGVD,
    author = "Avrorin, A. D. and others",
    collaboration = "Baikal-GVD",
    title = "{Search for High-Energy Neutrinos from GW170817 with the Baikal-GVD Neutrino Telescope}",
    eprint = "1810.10966",
    archivePrefix = "arXiv",
    primaryClass = "astro-ph.HE",
    doi = "10.1134/S0021364018240025",
    journal = "JETP Lett.",
    volume = "108",
    number = "12",
    pages = "787--790",
    year = "2018"
}

@article{ELOWEN_LVKO4followUp,
    author = "Lamoureux, Mathieu and others",
    collaboration = "IceCube",
    title = "{Probing neutrino emission at GeV energies from compact binary mergers detected during O1-O4a with the IceCube Neutrino Observatory}",
    eprint = "2507.13698",
    archivePrefix = "arXiv",
    primaryClass = "astro-ph.HE",
    reportNumber = "PoS-ICRC2025-947",
    doi = "10.22323/1.501.0947",
    journal = "PoS",
    volume = "ICRC2025",
    pages = "947",
    year = "2025"
}
\bibliographystyle{aasjournalv7}

\end{document}